\documentclass[10pt]{JHEP3}
\usepackage{epsfig,multicol,bbm}
\usepackage{epsfig}
\usepackage{amsmath}
\usepackage{epsfig}
\usepackage{psfrag}
\usepackage{amsfonts}
\usepackage{graphicx}
\usepackage{dcolumn}
\usepackage{bm}

\newcommand{\vect}{\left ( \begin{array}{c}}
\newcommand{\evect}{\end{array} \right )}
\newcommand{\fft}[2]{{\frac{#1}{#2}}}

\usepackage{epsfig}
\usepackage{psfrag}
\usepackage{graphicx}
\usepackage{dcolumn}
\usepackage{bm}
\def\fsl#1{\setbox0=\hbox{$#1$}                 
   \dimen0=\wd0                                 
   \setbox1=\hbox{/} \dimen1=\wd1               
   \ifdim\dimen0>\dimen1                        
      \rlap{\hbox to \dimen0{\hfil/\hfil}}      
      #1                                        
   \else                                        
      \rlap{\hbox to \dimen1{\hfil$#1$\hfil}}   
      /                                         
   \fi}                                         %

\title{
The Subleading Term of the Strong Coupling Expansion of the Heavy-Quark Potential
in a $\mathcal N=4$ Super Yang-Mills Vacuum}

\author{Shao-xia Chu$^{\dagger a}$,
Defu Hou $^{\dagger a}$,  Hai-cang Ren $^{\dagger b}$,$^{\dagger a}$\\

{$^{\dagger a}$ Institute of Particle Physics, Huazhong Normal
University, Wuhan 430079, China}\\
{E-mail:chusx@iopp.ccnu.edu.cn, ~hdf@iopp.ccnu.edu.cn}\\
{$^{\dagger b}$Physics Department, The Rockefeller University,
1230 York Avenue, New York, NY 10021-6399} \\
{E-mail:~ren@mail.rockefeller.edu}\\
 }

\abstract { Applying the AdS/CFT correspondence, the expansion of
the heavy-quark potential of ${\cal N}=4$ supersymmetric
Yang-Mills theory at large $N_c$ is carried out to the sub-leading
term in the large 't Hooft coupling at zero temperature. The
strong coupling corresponds to the semi-classical expansion of the
string-sigma model, the gravity dual of the Wilson loop operator,
with the sub-leading term expressed in terms of functional
determinants of fluctuations. The singularities of these
determinants are examined and their contributions are evaluated
numerically. }

\begin{document}
\section{Introduction}

AdS/CFT duality \cite{Maldacena:1997re,Gubser:1998bc,Witten:1998qj,MaldacenaReview}
remains an active field of research. Motivated by the isomorphism between
the isometry group of AdS$_5$ and the conformal group in four dimensions,
it was conjectured by Maldacena that a string theory in AdS$_5\times S^5$ corresponds
to a four dimensional conformal field theory on the boundary. A prominent
implication of the conjecture is the correspondence between the type IIB superstring theory formulated on
AdS$_5\times S^5$ and $\mathcal N=4$ supersymmetric Yang-Mills
theory (SYM) with the isometry group $O(6)$ of $S^5$ dual to the R-symmetriy group $SU(4)$
of SYM. In particular, the supergravity limit of the string theory corresponds
to the leading behavior of SYM at large $N_c$ and large 't Hooft coupling
\begin{equation}
\lambda \equiv g_{\rm YM}^2N_c = \fft{L^4}{\alpha'^2}.
\label{eq:ldef}
\end{equation}
with $L$ the AdS radius and $\alpha'$ the reciprocal of the string
tension. This relation thereby opens a new avenue to explore the
strong coupling properties of SYM and sheds new lights on strongly
coupled QGP created in RHIC in spite of the difference between SYM
and QCD. Among notable successes on the RHIC phenomenology are the
equation of state \cite{Witten:1998}, the viscosity
ratio\cite{Policastro:2001yc} and jet quenching parameters
\cite{Liu:2006ug} as well as the energy loss\cite{Herzog:2006gh}.

The heavy quark potential (the potential energy between a heavy
quark and its anti-particle) of QCD is an important quantity that
probes the confinement mechanism in the hadronic phase and the meson
melting in the plasma phase. It is extracted from the expectation of
a Wilson loop operator, which can be measured on a lattice. In the
case of $\mathcal N=4$ SYM, the AdS/CFT duality relates the Wilson
loop expectation value to the path integral of the string-sigma
action developed in Ref.\cite{Metsaev:1998it} for the worldsheet in
the AdS$_5\times S^5$ bulk spanned by the loop on the boundary. To
the leading order of strong coupling, the path integral is given by
its classical limit, which is the minimum area of the world sheet.
From the Wilson loop of a pair of parallel lines, Maldacena
extracted the potential function in $\mathcal N=4$ SYM at zero
temperature\cite{Maldacena:1998im},
\begin{equation}
V(r)=-\frac{4\pi^2}{\Gamma^4\left(\frac{1}{4}\right)}\frac{\sqrt{\lambda}}{r}
\simeq -0.2285\frac{\sqrt{\lambda}}{r}
\label{leading}
\end{equation}
with $r$ the distance between the quark and the antiquark.
Introducing a black hole in AdS bulk, the potential at nonzero
temperature as well as that for moving quarks have been obtained
by a number of authors\cite{Rey:1998bq}\cite{Liu2}. The field
theoretic aspects of the potential (\ref{leading}) and  its finite
temperature counterpart as well as their implications on RHIC
physics were discussed in
Ref.\cite{erickson}\cite{shuryak}\cite{Liu2}. As was pointed out
in Ref.\cite{Maldacena:1998im}, the "heavy quarks" underlying the
Wilson loop (\ref{leading}) in ${\cal N}=4$ SYM are actually heavy
W bosons resulted in a Higgs mechanism, which implement the
fundamental representation of $SU(N_c)$. Since the function
(\ref{leading}) measures the force between two static fundamental
color objects, we shall borrow the terminology of QCD by naming it
the heavy quark potential throughout this paper.

The strong coupling
expansion of the SYM Wilson loop corresponds to the semi-classical
expansion of the string-sigma action and reads
\begin{equation}
V(r)=-\frac{4\pi^2}{\Gamma^4\left(\frac{1}{4}\right)}\frac{\sqrt{\lambda}}{r}
\Big[1+\frac{\kappa}{\sqrt{\lambda}}+O\left(\frac{1}{\lambda}\right)\Big]
\label{expansion}
\end{equation}
for the heavy quark potential. Computing the coefficient $\kappa$ is
the main subject of the present paper. $\kappa$ comes from the one
loop effective action of the world sheet fluctuations around its
minimum area. This effective action has been obtained explicitly for
some simple Wilson loops including parallel
lines\cite{Forste:1999qn} \cite{Drukker:2000ep} and is expressed in
terms of functional determinants. Evaluating these determinants, we
end up with the numerical value of $\kappa$,
\begin{equation}
\kappa\simeq -1.33460.
\label{finalresult}
\end{equation}

The classical solution of the string-sigma model and the one loop
effective action underlying $\kappa$ is briefly reviewed in the
next section. There we also outline our strategy of computation,
which is along the line suggested in \cite{Drukker:2000ep}. We
parametrize the string world sheet of the single Wilson line or
parallel lines by conformal coordinates. Then a scaling
transformation is made that leaves the measure of the spectral
problem of the functional determinants trivial. Instead of solving
the eigenvalue problem of the operators underlying the
determinants, we use the method employed in
\cite{Kruczenski:2008zk}, which amounts to solve a set of ordinary
differential equations. Unlike the straight Wilson line and the
circular Wilson loop dealt with in \cite{Kruczenski:2008zk}, some
of differential equations for the parallel lines are not
analytically tractable. The presence of various singularities
makes numerical works highly nontrivial. It is critical to isolate
the singularities analytically in order to obtain a robust
numerical result. So we did and the procedure is described in
sections 3 and 4. The finite terms of the scaling transformation
of the determinants involved are examined in section 5 and we find
them adding up to zero. In section 6, we discuss our results along
with few open questions. Some technical details are explained in
appendices. Throughout the paper, we shall work with Euclidean
signature with the AdS radius $L$ set to one.


\section{The one-loop effective action}


Let us begin with a brief review of the classical limit that leads to the leading order
potential (\ref{leading}). The string-sigma action in this limit reduces to the
Nambu-Goto action
\begin{equation}
S_{\rm NG}=\frac{1}{2\pi\alpha^\prime}\int d^2\sigma\sqrt{g},
\label{NG}
\end{equation}
with $g$ the determinant of the induced metric on the string world sheet embedded in the target space, i.e.
\begin{equation}
g_{\alpha\beta}=G_{\mu\nu}\frac{\partial X^\mu}{\partial\sigma^\alpha}
\frac{\partial X^\nu}{\partial\sigma^\beta}
\end{equation}
where $X^\mu$ and $G_{\mu\nu}$ are the target space coordinates and the metric, and
$\sigma^\alpha$ with ($\alpha=0,1$) parametrize the world sheet. The target space here
is AdS$_5\times S^5$, whose metric may be written as
\begin{equation}
ds^2=\frac{1}{z^2}(dt^2+d\vec x^2+dz^2)+d\Omega_5^2 \label{target}
\end{equation}
with $d\Omega_5$ the element of the solid angle of S$^5$. The physical 3-brane resides
on the AdS boundary $z=0$. The string world sheets considered in this paper are all
projected onto a point of S$^5$ in the classical limit.

The Wilson loop of a static heavy quark, denoted by ${\cal C}_1$, is a straight line winding up the Euclidean time
periodically at the AdS boundary. The corresponding world sheet in the AdS bulk can be parametrized by $t$ and
$z$ with $\vec x$ constant and extends all the way to AdS horizon, $z\to\infty$.
The induced metric is that of AdS$_2$, given by
\begin{equation}
ds^2[{\cal C}_1]=\frac{1}{z^2}(dt^2+dz^2)
\label{1quark}
\end{equation}
with the scalar curvature
\begin{equation}
R=-2.
\label{r1}
\end{equation}
Substituting the metric (\ref{1quark}) into (\ref{NG}), we find the self-energy of
the heavy quark
\begin{equation}
E[{\cal C}_1]=\frac{1}{T}S_{\rm NG}[{\cal
C}_1]=\frac{1}{2\pi\alpha^\prime}\int_\delta^\infty\frac{dz}{z^2}.
\label{s1}
\end{equation}
with $T\to\infty$ the time period.
Notice that we have pulled the physical brane slightly off the boundary to the radial
coordinate $z=\delta$, as a regularization of the divergence pertaining the lower limit
of the integral (\ref{s1}).

The total energy of a pair of a heavy quark and a heavy antiquark
separated by a distance $r$, can be extracted from the Wilson loop
consisting of two parallel lines each winding up the Euclidean
time at the boundary. This Wilson loop will be denoted by ${\cal
C}_2$ and the world sheet in the bulk can be parametrized by $t$
and $z$ with $x^1=\xi(z)$ and $x^2, x^3={\rm const.}$. The
function $\xi(z)$ is determined by substituting the induced metric
\begin{equation}
ds^2[{\cal C}_2]=\frac{1}{z^2}\{dt^2+\Big[\left(\frac{d\xi}{dz}\right)^2+1\Big]dz^2\},
\label{s2}
\end{equation}
into the action (\ref{NG}) and minimizing it. We have
\begin{equation}
\xi=\pm\int_z^{z_0}dz'\frac{z'^2}{\sqrt{z_0^4-z'^4}}.
\label{xi}
\end{equation}
The maximum bulk extension of the world sheet, $z_0$, is determined by the distance $r$
between the two lines at the boundary and we find that
\begin{equation}
z_0=\frac{\Gamma^2\left(\frac{1}{4}\right)}{(2\pi)^{\frac{3}{2}}}r
\label{z0}
\end{equation}
Substituting (\ref{xi}) into ({\ref{s2}), we end up with the induced metric
\begin{equation}
ds^2[{\cal C}_2]=\frac{1}{z^2}\left(dt^2+\frac{z_0^4}{z_0^4-z^4}dz^2\right),
\label{ss2}
\end{equation}
and the scalar curvature
\begin{equation}
R=-2\left(1+\frac{z^4}{z_0^4}\right)
\label{r2}
\end{equation}
The energy of the heavy quark pair is therefore given by,
\begin{equation}
E[{\cal C}_2]=\frac{1}{T}S_{\rm NG}[{\cal
C}_2]=\frac{1}{\pi\alpha^\prime}z_0^2\int_\delta^{z_0}\frac{dz}{z^2\sqrt{z_0^4-z^4}},
\label{e2}
\end{equation}
where the same regularization is applied to the lower limit of the integral.

The heavy quark potential is obtained by subtracting from (\ref{e2}) the self energy of each
quark(antiquark), i.e.
\begin{equation}
V=\lim_{\delta\to 0^+}(E[{\cal C}_2]-2E[{\cal C}_1])=\frac{1}{\pi\alpha^\prime}\Big[\int_0^{z_0}dz
\left(\frac{z_0^2}{z^2\sqrt{z_0^4-z^4}}-\frac{1}{z^2}\right)-\frac{1}{z_0}\Big],
\end{equation}
and is divergence free. Carrying out the integral and substituting in the relations
(\ref{z0}), we derive (\ref{leading}).

The one loop effective action, $W$ is obtained by expanding the string-sigma action
of Ref.\cite{Metsaev:1998it} to the
quadratic order of the fluctuating coordinates around the minimum area and carrying out
the path integral\cite{Forste:1999qn}\cite{Drukker:2000ep}. We have
\begin{equation}
W[{\cal C}_1]=-\ln\Big[\frac{{\det}^4(-i\gamma^\alpha\nabla_\alpha+\tau_3)}
{{\det}^{\frac{3}{2}}(-\nabla^2+2){\det}^{\frac{5}{2}}(-\nabla^2)}\Big],
\label{straight}
\end{equation}
for the static quark or antiquark and
\begin{equation}
W[{\cal C}_2]=-\ln\Big[\frac{{\rm det}^4(-i\gamma^\alpha\nabla_\alpha+\tau_3)}
{{\det}^{\frac{1}{2}}(-\nabla^2+4+R)\,
{\det}(-\nabla^2+2)\,{\det}^{\frac{5}{2}}(-\nabla^2)}\Big],
\label{parallel}
\end{equation}
for the quark pair. The determinants in the denominators of
(\ref{straight}) and (\ref{parallel}) come from the fluctuations
of three transverse coordinates of the AdS sector and five
coordinates of $S^5$ with the Laplacian given by the metric
(\ref{1quark}) or (\ref{ss2}). The determinants in the numerators
come from the fermionic fluctuations, where we have introduced 2d
gamma matrices, $\gamma_0=\gamma^0=\tau_2$,
$\gamma_1=\gamma^1=\tau_1$ and $\gamma_0\gamma_1=-i\tau_3$ with
$\tau_1$, $\tau_2$ and $\tau_3$ the three Pauli matrices. In terms
of the zweibein of the world sheet, $e_\alpha^j$, we have
$\gamma_\alpha\equiv e_\alpha^j\gamma_j$ with $j=0,1$ and the
covariant derivative
\begin{equation}
\nabla_\alpha=\frac{\partial}{\partial \sigma^\alpha}+\frac{1}{8}
[\gamma_i,\gamma_j]\omega_\alpha^{ij}
\end{equation}
with $\omega_\alpha^{ij}$ the spin connection corresponding to (\ref{1quark}) or (\ref{ss2}). The power
"4" comes from eight 2d Majorana fermions each of which contributes a power 1/2.
The one loop correction to the heavy quark potential is then
\begin{equation}
\Delta V=\lim_{T\to\infty}\frac{1}{T}\lim_{\delta\to 0^+}(W[{\cal C}_2]-2W[{\cal C}_1]).
\label{oneloop}
\end{equation}

The effective action $W[{\cal C}_1]$ or $W[{\cal C}_2]$ suffers from the usual logarithmic
UV divergence, which is proportional to the volume part of the Euler character
\begin{equation}
\int_{z>\delta}dt dz\sqrt{g}R
\label{euler}
\end{equation}
of each world sheet with the same coefficient of
proportionality\cite{Drukker:2000ep}. It follows from (\ref{1quark}), (\ref{r1}),
(\ref{ss2}) and (\ref{r2}) that the integral
(\ref{euler}) for the parallel lines is exactly twice of that for
the single line in the limit $\delta\to 0$. We have indeed that
\begin{equation}
\int d^2\sigma\sqrt{g}R=T\int_\delta^\infty\frac{dz}{z^2}(-2)=-\frac{2T}{\delta}
\end{equation}
for the single line and
\begin{equation}
\int
d^2\sigma\sqrt{g}R=2T\int_\delta^{z_0}dz\frac{z_0^2}{\sqrt{z_0^4-z^4}}
(-2)\left(1+\frac{z^4}{z_0^4}\right)=\left.
\frac{4T}{z}\sqrt{1-\frac{z^4}{z_0}}\right |_\delta^{z_0}
=-\frac{4T}{\delta}+O(\delta^3).
\end{equation}
for the parallel lines. Therefore the UV divergence as well as the
conformal anomaly cancel in the combination of (\ref{oneloop}) in
the limit $\delta\to 0$. As a contrast, the volume integral $\int
d^2\sigma\sqrt{g}$ of the parallel lines differs from twice of
that of a straight line by a finite quantity in the same limit.
The UV divergence associated to the volume integral cancels within
each effective action of (\ref{straight}) and
(\ref{parallel})\cite{fn1}. Furthermore the limit $\delta\to 0^+$
of the UV finite term of (\ref{oneloop}) also exists as we shall
see.

The world sheet of the parallel lines covers the coordinate patch
$(t,z)$ twice, which gives rise to an artificial singularity of
the Laplacian's in (\ref{parallel}) at $z=z_0$ and adds
difficulties to the numerical works. To avoid the problem, we
shall work with a conformal coordinate patch $(\tau,\sigma)$ that
the world sheet (\ref{ss2}) covers only once. This is also
suggested in \cite{Drukker:2000ep}. The new coordinates involve
Jacobi elliptic functions \cite{Whittaker}\cite{Wang}of modulo
$k=\frac{1}{\sqrt{2}}$ and are defined by
\begin{equation}
z=z_0{\rm cn}\sigma \qquad t=\frac{z_0}{\sqrt{2}}\tau
\end{equation}
In terms of the new coordinates, the metric (\ref{ss2}) takes the
form
\begin{equation}
ds^2[{\cal C}_2]=\frac{1}{2{\rm cn}^2\sigma}(d\tau^2+d\sigma^2),
\label{conformal}
\end{equation}
and the scalar curvature (\ref{r2}) becomes
\begin{equation}
R=-2(1+{\rm cn}^4\sigma).
\end{equation}
The nonzero component of the spin connection with cartesian
indexes (0,1) referring to the coordinate differentials $d\tau$
and $d\sigma$ reads
\begin{equation}
\omega_\tau^{01}=-\omega_\tau^{10}=\frac{{\rm sn}\sigma{\rm dn}\sigma}{{\rm cn}\sigma}.
\label{spin}
\end{equation}
We shall use the the same time variable $\tau$ to describe the world sheet of the straight
line and rescale the $z$ coordinate by $z=\frac{z_0}{\sqrt{2}}\zeta$, leaving the
conformal structure of (\ref{1quark}) intact, i.e.
\begin{equation}
ds^2[{\cal C}_1]=\frac{1}{\zeta^2}(d\tau^2+d\zeta^2).
\label{conformal1}
\end{equation}
The spin connection corresponding to (\ref{spin}) is given by
$\omega_\tau^{01}=-\omega_\tau^{10}=-\frac{1}{\zeta}$. The range of each coordinate
variable is $-\frac{{\cal T}}{2}\le\tau\le\frac{{\cal T}}{2}$,
$-K\le\sigma\le K$ and $0\le\zeta<\infty$ where ${\cal
T}=\frac{\sqrt{2}}{z_0}T$ and $K$ is the complete elliptic integral
of the first kind,
\begin{equation}
K=\frac{\Gamma^2\left(\frac{1}{4}\right)}{4\sqrt{\pi}}\simeq 1.8541.
\end{equation}
The operators underlying the determinants of (\ref{straight}) are given explicitly by
\begin{equation}
\Delta_0[{\cal C}_1]\equiv-\nabla^2=-\zeta^2\left(\frac{\partial^2}{\partial\tau^2}
+\frac{\partial^2}{\partial\zeta^2}\right)\equiv\zeta^2\hat\Delta_0[{\cal C}_1]
\label{Delta01}
\end{equation}
\begin{equation}
\Delta_1[{\cal C}_1]\equiv-\nabla^2+2=-\zeta^2\left(\frac{\partial^2}{\partial\tau^2}
+\frac{\partial^2}{\partial\zeta^2}\right)+2\equiv\zeta^2\hat\Delta_1[{\cal C}_1]
\label{Delta11}
\end{equation}
and
\begin{equation}
D_F[{\cal C}_1]\equiv-i\gamma^\alpha\nabla_\alpha+\tau_3=-i\zeta\left(\frac{d}{d\zeta}-\frac{1}{2\zeta}\right)\tau_1
-i\zeta\frac{\partial}{\partial\tau}\tau_2+\tau_3
\equiv\zeta\hat D_F[{\cal C}_1].
\label{DF1}
\end{equation}
Similarly, the explicit expressions of the operators underlying the
determinants of (\ref{parallel}) reads
\begin{equation}
\Delta_0[{\cal C}_2]\equiv-\nabla^2=-2{\rm cn}^2\sigma\left(\frac{\partial^2}{\partial\tau^2}
+\frac{\partial^2}{\partial\sigma^2}\right)\equiv 2{\rm cn}^2\sigma\hat\Delta_0[{\cal C}_2],
\label{Delta02}
\end{equation}
\begin{equation}
\Delta_1[{\cal C}_2]\equiv-\nabla^2+2=-2{\rm cn}^2\sigma\left(\frac{\partial^2}{\partial\tau^2}
+\frac{\partial^2}{\partial\sigma^2}\right)+2\equiv 2{\rm cn}^2\sigma\hat\Delta_1[{\cal C}_2],
\label{Delta12}
\end{equation}
\begin{equation}
\Delta_2[{\cal C}_2]\equiv-\nabla^2+4+R=-2{\rm cn}^2\sigma\left(\frac{\partial^2}{\partial\tau^2}
+\frac{\partial^2}{\partial\sigma^2}\right)+2(1-{\rm cn}^4\sigma)\equiv 2{\rm cn}^2\sigma\hat\Delta_2[{\cal C}_2],
\label{Delta22}
\end{equation}
and
\begin{equation}
D_F[{\cal C}_2]\equiv-i\gamma^\alpha\nabla_\alpha+\tau_3
=-i\sqrt{2}{\rm
cn}\sigma\left(\frac{\partial}{\partial\sigma}+\frac{{\rm
sn}\sigma{\rm dn}\sigma}{2{\rm cn}\sigma}\right)\tau_1
-i\sqrt{2}{\rm cn}\sigma\frac{\partial}{\partial\tau}\tau_2+\tau_3
\equiv{\rm cn}\sigma\hat D_F[{\cal C}_2]. \label{DF2}
\end{equation}
The difference between the operators with hats and those without
hats is the measure of the spectral problem defined by them. While
the measure is trivial with respect to the operators with hats,
changing the measure may introduce additional terms to the
logarithm of each determinant and their contribution will be
examined in section V. For this reason, the effective action is
decomposed into two pieces, i.e. $W[{\cal C}_1]= W_1[{\cal
C}_1]+W_2[{\cal C}_1]$ for the single Wilson line and $W[{\cal
C}_2]= W_1[{\cal C}_2]+W_2[{\cal C}_2]$ for the parallel lines. We
define
\begin{equation}
W_1[{\cal C}_1]=-\ln\frac{\det^4\hat D_F[{\cal C}_1]}
{\det^{\frac{5}{2}}\hat\Delta_0[{\cal C}_1]\det^{\frac{3}{2}}\hat\Delta_1[{\cal C}_1]},
\end{equation}
\begin{equation}
W_2[{\cal C}_1]=-4\ln\frac{|\det D_F[{\cal C}_1]|}{|\det\hat
D_F[{\cal C}_1]|} +\frac{5}{2}\ln\frac{\Delta_0[{\cal
C}_1]}{\hat\Delta_0[{\cal C}_1]}
+\frac{3}{2}\ln\frac{\Delta_1[{\cal C}_1]}{\hat\Delta_1[{\cal
C}_1]}, \label{DeltaS21}
\end{equation}
\begin{equation}
W_1[{\cal C}_2]=-\ln\frac{\det^4\hat D_F[{\cal C}_2]}
{\det^{\frac{5}{2}}\hat\Delta_0[{\cal C}_2]\det\hat\Delta_1[{\cal C}_2]\det^{\frac{1}{2}}\hat\Delta_2[{\cal C}_2]},
\end{equation}
and
\begin{equation}
W_2[{\cal C}_2]=-4\ln\frac{|\det D_F[{\cal C}_2]|}{|\det\hat
D_F[{\cal C}_2]|} +\frac{5}{2}\ln\frac{\Delta_0[{\cal
C}_2]}{\hat\Delta_0[{\cal C}_2]} +\ln\frac{\Delta_1[{\cal
C}_2]}{\hat\Delta_1[{\cal C}_2]}
+\frac{1}{2}\ln\frac{\Delta_2[{\cal C}_2]}{\hat\Delta_2[{\cal
C}_2]}, \label{DeltaS22}
\end{equation}
Correspondingly, the coefficient $\kappa$ defined in (\ref{expansion}) is given by
$\kappa=\kappa_1+\kappa_2$ with
\begin{equation}
\kappa_1\equiv\frac{\Gamma^2\left(\frac{1}{4}\right)}{\sqrt{\pi}{\cal T}}
\lim_{\delta\to 0^+}(W_1[{\cal C}_2]-2W_1[{\cal C}_1]),
\label{kappa1def}
\end{equation}
and
\begin{equation}
\kappa_2\equiv\frac{\Gamma^2\left(\frac{1}{4}\right)}{\sqrt{\pi}{\cal T}}
\lim_{\delta\to 0^+}(W_2[{\cal C}_2]-2W_2[{\cal C}_1]),
\label{kappa2def}
\end{equation}
where we have used the relation between $T$ and ${\cal T}$ and
converted $z_0$ to $r$ via (\ref{z0}).

Making a Fourier transformation of the time variable $\tau$, each
functional determinant of (\ref{straight}) and (\ref{parallel}) is
factorized as an infinite product of its Fourier components with
each Fourier component obtained by replacing the time derivative
$\frac{\partial}{\partial\tau}$ in $\hat\Delta$'s of
(\ref{Delta01})-(\ref{DF2}) by $-i\omega$ with $\omega$ a
frequency variable. Substituting the Fourier product of
(\ref{straight}) and that of (\ref{parallel}) into
(\ref{kappa1def}), we find that

\begin{equation}
\kappa_1=\frac{\Gamma^2\left(\frac{1}{4}\right)}{\sqrt{\pi}}
\int_{-\infty}^\infty\frac{d\omega}{2\pi}\ln\frac{{\cal
R}_2(\omega)}{{\cal R}_1^2(\omega)}
=\frac{\Gamma^2\left(\frac{1}{4}\right)}{\pi^{\frac{3}{2}}}
\int_0^\infty d\omega\ln\frac{{\cal R}_2(\omega)}{{\cal
R}_1^2(\omega)} \label{coeff}
\end{equation}

The functions ${\cal R}_1(\omega)$ and ${\cal R}_2(\omega)$ are the
Fourier components of the determinant ratios of (\ref{straight}) and
(\ref{parallel}), given by
\begin{equation}
{\cal R}_1(\omega)=\frac{{\rm det}{\cal D}_+^2(\omega){\rm det}{\cal D}_-^2(\omega)}
{{\rm det}{\cal D}_0^{\frac{5}{2}}(\omega){\rm det}{\cal D}_1^{\frac{3}{2}}(\omega)}
\end{equation}
and
\begin{equation}
{\cal R}_2(\omega)=\frac{{\rm det}D_+^2(\omega){\rm det}D_-^2(-\omega)}
{{\rm det}D_0^{\frac{5}{2}}(\omega){\rm det}D_1(\omega){\rm det}D_2^{\frac{1}{2}}(\omega)},
\end{equation}
where the Fourier transformation of the operators $\hat\Delta$'s and $\hat D_F$'s are given by
\begin{equation}
{\cal D}_0(\omega)={\cal D}_-(\omega)=-\frac{d^2}{d\zeta^2}+\omega^2,
\label{straightd0}
\end{equation}
\begin{equation}
{\cal D}_1(\omega)={\cal D}_+(\omega)=-\frac{d^2}{d\zeta^2}+\omega^2+\frac{2}{\zeta^2},
\label{straightd1}
\end{equation}
\begin{equation}
D_0(\omega)=-\frac{d^2}{d\sigma^2}+\omega^2,
\label{paralleld0}
\end{equation}
\begin{equation}
D_1(\omega)=-\frac{d^2}{d\sigma^2}+\omega^2+\frac{1}{{\rm cn}^2\sigma},
\label{paralleld1}
\end{equation}
\begin{equation}
D_2(\omega)=-\frac{d^2}{d\sigma^2}+\omega^2+\frac{1}{{\rm cn}^2\sigma}-{\rm cn}^2\sigma,
\label{paralleld2}
\end{equation}
and
\begin{equation}
D_\pm(\omega)=-\frac{d^2}{d\sigma^2}+\left(\omega^2+\frac{1\mp\sqrt{2}{\rm
sn}\sigma{\rm dn}\sigma} {2{\rm cn}^2\sigma}\right).
\label{paralleldf}
\end{equation}
Let us explain the transformation we made on the fermionic
determinants $\det\hat D_F[{\cal C}_1]$ and $\det\hat D_F[{\cal
C}_2]$. Replacing the time derivatives in (\ref{DF1}) and
(\ref{DF2}) by $-i\omega$, we find that
\begin{equation}
\hat D_F[{\cal C}_1]=-i\left(\frac{d}{d\zeta}-\frac{1}{2\zeta}\right)\tau_1-\omega\tau_2+\frac{1}{\zeta}\tau_3
\end{equation}
and
\begin{equation}
\hat D_F[{\cal C}_2]=-i\left(\frac{d}{d\sigma}+\frac{{\rm sn}\sigma{\rm dn}\sigma}{2{\rm cn}\sigma}\right)\tau_1
-\omega\tau_2+\frac{1}{\sqrt{2}{\rm cn}\sigma}\tau_3.
\end{equation}
It is straightforward to verify that
\begin{equation}
\hat D_F^2[{\cal C}_1]=\sqrt{\zeta}U{\rm diag.}({\cal D}_+(\omega),{\cal D}_-(\omega))U^\dagger\frac{1}{\sqrt{\zeta}}
\end{equation}
and
\begin{equation}
\hat D_F^2[{\cal C}_2]=\sqrt{{\rm cn}\sigma}U{\rm diag.}(D_+(\omega),D_-(\omega))U^\dagger\frac{1}{\sqrt{{\rm cn}\sigma}},
\end{equation}
where $U$  is a $2\times 2$ matrix that diagonalizes $\tau_2$, ${\cal D}_\pm(\omega)$ and $D_\pm(\omega)$ are
given above by (\ref{straightd0}) (\ref{straightd1}) and (\ref{paralleldf}). Therefore
$\det^4\hat D_F[{\cal C}_1]=\det^2{\cal D}_+(\omega)\det^2{\cal D}_-(\omega)$ and
$\det^4\hat D_F[{\cal C}_2]=\det^2D_+(\omega)\det^2D_-(\omega)$.

The evaluation of the integral (\ref{coeff}) will be discussed in the next two sections.

\section{The evaluation of $\kappa_1$ ---- analytical part}

Evaluating a functional determinant stemming from a semi-classical
approximation to a quantum mechanical system is an old subject of
many research works. In one dimension a short cut was discovered by
a number of authors \cite{Gelfand}\cite{Kleinert} that does not require a solution
to the spectrum problem involved. Consider two functional operators
\begin{equation}
H_\alpha = -\frac{d^2}{dx^2}+V_\alpha(x)
\end{equation}
with $\alpha=1,2$ defined in the domain $a\le x\le b$ where
$V_\alpha(x)\ge 0$ under the Dirichlet boundary condition, it was
shown that the determinant ratio
\begin{equation}
\frac{{\rm det}H_2}{{\rm det}H_1}=\frac{f_2(b|a)}{f_1(b|a)}
\label{dratio}
\end{equation}
where $f_\alpha(x|a)$ is the solution of the homogeneous equation
\begin{equation}
H_\alpha\phi=0,
\label{homo}
\end{equation}
subject to the conditions $f_\alpha(0|a)=0$ and
$f_\alpha^\prime(0|a)=1$. In terms of a pair of linearly independent
solutions of (\ref{homo}), $(u_\alpha,v_\alpha)$,
\begin{equation}
f_\alpha(b|a)=\frac{u_\alpha(a)v_\alpha(b)-u_\alpha(b)v_\alpha(a)}
{W[u_\alpha,v_\alpha]}.
\label{lambda}
\end{equation}
where the Wronskian $W[u_\alpha,v_\alpha]$ is $x$-independent.
With appropriate modification of the conditions imposed on
$f_\alpha(x|a)$, the formula (\ref{dratio}) can be tailored to
cover other boundary conditions. This method has been employed
recently in \cite{Kruczenski:2008zk} to calculate the one loop
effective action of the single line ${\cal C}_1$ or that of a
circular Wilson loop. See \cite{Dunne} for a review on other
applications.

Coming back to the semi-classical correction of the heavy quark
potential, the operator $H_\alpha$ corresponds to one of the
operators (\ref{straightd0}) -(\ref{paralleldf}). We shall retain
$(u,v)$ for a pair of linearly independent solutions of the
homogeneous equation (\ref{homo}) with $H_\alpha$ given by an
operator pertaining to the single line and denote that of the
corresponding equation of the parallel lines by $(\eta,\xi)$. Eq.
(\ref{homo}) with $H_\alpha$ given by an operator of
(\ref{straightd0})-(\ref{paralleld0}) can be
solved analytically and we may choose the following pairs of
independent solutions
\begin{equation}
u_0=\sinh\omega\zeta\equiv u_0(\omega\zeta) \qquad v_0=e^{-\omega\zeta}\equiv v_0(\omega\zeta),
\label{uv0}
\end{equation}
\begin{equation}
u_1=\cosh\omega\zeta-\frac{\sinh\omega\zeta}{\omega\zeta}\equiv u_1(\omega\zeta)
\qquad
v_1=\left(1+\frac{1}{\omega\zeta}\right)e^{-\omega\zeta}\equiv v_1(\omega\zeta),
\label{uv1}
\end{equation}
\begin{equation}
u_+=u_1(\omega\zeta)
\qquad
v_+=v_1(\omega\zeta)
\label{uv+}
\end{equation}
\begin{equation}
u_-=u_0(\omega\zeta)
\qquad
v_-=v_0(\omega\zeta)
\label{uv-}
\end{equation}
and
\begin{equation}
\eta_0=\sinh\omega\sigma \qquad \xi_0=\cosh\omega\sigma.
\label{etaxi0}
\end{equation}
with their Wronskian's all given by
\begin{equation}
W[u_0,v_0]=W[\eta_0,\xi_0]=W[u_1,v_1]=W[u_\pm,v_\pm]=-\omega,
\end{equation}

The equations (\ref{homo}) with $H_\alpha$ given by (\ref{paralleld1}),
(\ref{paralleld2}) and (\ref{paralleldf}),
\begin{equation}
D_1(\omega)\phi=-\frac{d^2\phi}{d\sigma^2}+\left(\omega^2+\frac{1}{{\rm cn}^2\sigma}\right)\phi=0,
\label{potential1}
\end{equation}
\begin{equation}
D_2(\omega)\phi=-\frac{d^2\phi}{d\sigma^2}+\left(\omega^2+\frac{1}{{\rm
cn}^2\sigma} -{\rm cn}^2\sigma\right)\phi=0, \label{potential2}
\end{equation}
and
\begin{equation}
D_\pm(\omega)\phi=-\frac{d^2\phi}{d\sigma^2}+\left(\omega^2+\frac{1\mp\sqrt{2}{\rm sn}\sigma{\rm dn}\sigma}
{2{\rm cn}^2\sigma}\right)\phi=0.
\label{potential+}
\end{equation}
do not admit analytical solutions for $\omega\neq 0$.
Eqs.(\ref{potential1}) and (\ref{potential2}) have $\sigma=\pm K$
as regular points with the same pair of indexes (2,-1) there. The
equation $D_\pm(\omega)\phi=0$ has a regular point $\sigma=\mp K$
with the indexes (2,-1) and $\sigma=\pm K$ is an ordinary point of
it. We associate $\eta's$ to the vanishing solution at $\sigma=-K$
and $\xi's$ to the vanishing solution at $\sigma=K$ with the
normalization conditions
\begin{equation}
\lim_{\sigma\to -K}\frac{\eta_1(\sigma)}{\omega^2(\sigma+K)^2}
=\lim_{\sigma\to -K}\frac{\eta_2(\sigma)}{\omega^2(\sigma+K)^2}
=\lim_{\sigma\to -K}\frac{\eta_+(\sigma)}{\omega^2(\sigma+K)^2}=1
\label{etadef}
\end{equation}
and
\begin{equation}
\lim_{\sigma\to K}\frac{\xi_1(\sigma)}{\omega^2(K-\sigma)^2}
=\lim_{\sigma\to K}\frac{\xi_2(\sigma)}{\omega^2(K-\sigma)^2}
=\lim_{\sigma\to K}\frac{\xi_-(\sigma)}{\omega^2(K-\sigma)^2}=1.
\label{xidef}
\end{equation}
Furthermore, we require
\begin{equation}
\eta_-(-K)=\xi_+(K)=0
\label{cond1}
\end{equation}
and
\begin{equation}
\eta_-^\prime(-K)=-\xi_+^\prime(K)=\omega
\label{cond2}
\end{equation}
with the prime the derivative with respect $\sigma$.
On account of the eveness of $D_1(\omega)$ and $D_2(\omega)$ with respect to $\sigma$,
we have
\begin{equation}
\xi_{1,2}(\sigma)=\eta_{1,2}(-\sigma).
\label{symmetry1}
\end{equation}
It follows from the relation between $D_+(\omega)$ and $D_-(\omega)$ that
\begin{equation}
\eta_-(\sigma)=\xi_+(-\sigma) \qquad \xi_-(\sigma)=\eta_+(-\sigma)
\label{symmetry2}
\end{equation}
Each differential equation of (\ref{potential1}),
(\ref{potential2}) and (\ref{potential+}) is of the form of an one
dimensional Schroedinger equation in a non negative potential at
zero energy and does not admit a bound state subject to the
Dirichlet boundary condition. Therefore we expect that
\begin{equation}
\eta_{1,2}(\sigma)=\frac{C_{1,2}(\omega)}{\omega(K-\sigma)}+...,
\end{equation}
\begin{equation}
\eta_-(\sigma)=\frac{C_-(\omega)}{\omega(K-\sigma)}+...,
\end{equation}
as $\sigma\to K$ and
\begin{equation}
\xi_{1,2}(\sigma)=\frac{C_{1,2}(\omega)}{\omega(K+\sigma)}+...,
\end{equation}
\begin{equation}
\xi_+(\sigma)=\frac{C_+(\omega)}{\omega(K+\sigma)}+....
\end{equation}
as $\sigma\to -K$. The coefficients of divergence, $C_1(\omega)$,
$C_2(\omega)$ and $C_\pm(\omega)$ are related to the Wronskian's via
\begin{equation}
C_j(\omega)=-\frac{W[\eta_j,\xi_j]}{3\omega}
\label{CtoW}
\end{equation}
with $j=1,2,\pm$. We have, in addition,
\begin{equation}
\eta_+(K)=-\frac{W[\eta_+,\xi_+]}{\omega}
\qquad
\xi_-(-K)=-\frac{W[\eta_-,\xi_-]}{\omega}.
\label{alternative}
\end{equation}
It follows from the symmetry property (\ref{symmetry2}) that
\begin{equation}
C_+(\omega)=C_-(\omega)\equiv C_3(\omega).
\end{equation}

For $\omega>>1$, the solutions $\eta$'s and $\xi$'s can be approximated by WKB
method and we find the asymptotic forms
\begin{equation}
C_1(\omega)=\frac{3}{2}e^{2K\omega}\left(1-\frac{c}{\omega}+...\right),
\label{largeomega1}
\end{equation}
\begin{equation}
C_2(\omega)=\frac{3}{2}e^{2K\omega}\left(1-\frac{2c}{\omega}+...\right),
\label{largeomega2}
\end{equation}
and
\begin{equation}
C_3(\omega)=\frac{1}{2}e^{2K\omega}\left(1-\frac{c}{2\omega}+...\right),
\label{largeomega3}
\end{equation}
where the constant
\begin{equation}
c=\frac{2\pi^{\frac{3}{2}}}{\Gamma^2\left(\frac{1}{4}\right)}\simeq 0.84721
\end{equation}
The details of the derivation are shown in the appendix A. The small $\omega$
behavior can be obtained by introducing an alternative set of solutions, normalized differently,

\begin{equation}
\bar\eta_{1,2,+}(\sigma)\equiv\frac{\eta_{1,2,+}(\sigma)}{\omega^2}, \qquad
\bar\xi_{1,2,-}(\sigma)\equiv\frac{\xi_{1,2,-}(\sigma)}{\omega^2}
\end{equation}
and
\begin{equation}
\bar\eta_-(\sigma)\equiv\frac{\eta_-(\sigma)}{\omega}
\qquad \bar\xi_+(\sigma)\equiv\frac{\xi_+(\sigma)}{\omega}.
\end{equation}
Defining the coefficients
$\bar C$'s by the diverging behavior
\begin{equation}
\bar\eta_{1,2,-}(\sigma)=\frac{\bar C_{1,2,-}(\omega)}{K-\sigma}+...,
\end{equation}
as $\sigma\to K$ and
\begin{equation}
\bar\xi_{1,2,+}(\sigma)=\frac{\bar C_{1,2,+}(\omega)}{K+\sigma}+...,
\end{equation}
as $\sigma\to-K$, we find that $C_{1,2}(\omega)=\omega^3\bar C_{1,2}(\omega)$ and
$C_3(\omega)=C_\pm(\omega)=\omega^2\bar C_\pm(\omega)$. Since $\bar C_{1,2}(0),\bar C_\pm(0)\neq 0$
and are well defined (determined by eqs.(\ref{potential1})-(\ref{potential+})
at $\omega=0$, see Appendix B for details) we have the small $\omega$ behavior,
\begin{equation}
C_1(\omega)\simeq\frac{24\pi^{\frac{3}{2}}}{\Gamma^2\left(\frac{1}{4}\right)}\omega^3\simeq10.166557\omega^3,
\label{barc1}
\end{equation}
\begin{equation}
C_2(\omega)\simeq\frac{12\pi^{\frac{3}{2}}}{\Gamma^2\left(\frac{1}{4}\right)}\omega^3\simeq5.0832785\omega^3,
\label{barc2}
\end{equation}
and
\begin{equation}
C_3(\omega)\simeq 4\omega^2,
\label{barc3}
\end{equation}

In the regularized version, the physical brane, located at
$z=\delta$ cut the world sheet of the parallel lines at
$-K+\epsilon$ and $K-\epsilon$ with
\begin{equation}
\delta=z_0{\rm cn}(K-\epsilon)\simeq\frac{z_0}{\sqrt{2}}\epsilon.
\label{regparrelation}
\end{equation}
In another word, the domain of $\sigma$ coordinate is
$[-K+\epsilon,K-\epsilon]$ under the regularization, and we shall
impose the Dirichlet boundary condition there. The corresponding
domain of the single line becomes $\epsilon<\zeta<Z$ with $Z$ a
large $\zeta$ cutoff which will be set to infinity at the end.
Designate ${\cal U}_\alpha(\omega)$ to the quantity (\ref{lambda})
of the single Wilson line and $U_\alpha(\omega)$ to that of the
parallel lines with $\alpha=0,1,2,\pm$ corresponding to the
indexes of the operators (\ref{straightd0})-(\ref{paralleldf}), we
have \cite{fn2}

\begin{equation}
{\cal R}_1(\omega)=\frac{{\cal U}_+^2(\omega){\cal U}_-^2(\omega)}
{{\cal U}_0^{\frac{5}{2}}(\omega){\cal D}_1^{\frac{3}{2}}(\omega)}
=\left(1+\frac{1}{\omega\epsilon}\right)^{\frac{1}{2}}.
\label{finalR1}
\end{equation}
and
\begin{equation}
{\cal R}_2(\omega)=\frac{U_+^2(\omega)U_-^2(\omega)}
{U_0^{\frac{5}{2}}(\omega)U_1(\omega)U_2^{\frac{1}{2}}(\omega)}.
\label{R2}
\end{equation}
The last step of (\ref{finalR1}) follows from the solutions (\ref{uv0})-(\ref{uv-}),
which imply that
\begin{equation}
{\cal U}_0(\omega)={\cal U}_-(\omega)=\frac{1}{2\omega}e^{\omega(Z-\epsilon)}
\end{equation}
and
\begin{equation}
{\cal U}_1(\omega)={\cal U}_+(\omega)=\frac{1}{2\omega}
\left(1+\frac{1}{\omega\epsilon}\right)e^{\omega(Z-\epsilon)}
\end{equation}
as $Z\to\infty$. It follows from (\ref{etaxi0}) that
\begin{equation}
U_0(\omega)=\frac{\sinh2(K-\epsilon)\omega}{\omega}.
\label{U0}
\end{equation}
The symmetry (\ref{symmetry2}) implies that
$U_+(\omega)=U_-(\omega)$.

To proceed, let us introduce $\omega_0$
that satisfies the inequality $1<<\omega_0<<\frac{1}{\epsilon}$ and
divide the integral (\ref{coeff}) into two terms,
$\kappa_1=\kappa_<+\kappa_>$, with
\begin{equation}
\kappa_<=\frac{\Gamma^2\left(\frac{1}{4}\right)}{\pi^{\frac{3}{2}}}
\int_0^{\omega_0}d\omega\ln\frac{{\cal R}_2(\omega)}{{\cal R}_1^2(\omega)}
\label{lower}
\end{equation}
and
\begin{equation}
\kappa_>=\frac{\Gamma^2\left(\frac{1}{4}\right)}{\pi^{\frac{3}{2}}}
\int_{\omega_0}^\infty d\omega\ln\frac{{\cal R}_2(\omega)}{{\cal R}_1^2(\omega)}.
\label{upper}
\end{equation}
For the integrand of (\ref{lower}), we may approximate
\begin{equation}
{\cal R}_1(\omega)\simeq\frac{1}{\sqrt{\omega\epsilon}},
\label{approx1}
\end{equation}
\begin{equation}
U_{1,2}(\omega)\simeq\frac{C_{1,2}(\omega)}{3\omega^3\epsilon^2}
\label{approx2}
\end{equation}
and
\begin{equation}
U_\pm(\omega)\simeq\frac{C_3(\omega)}{\omega^2\epsilon}.
\label{approx3}
\end{equation}
Only one term of the numerator of (\ref{lambda}) contributes to each case of
(\ref{approx2}) and (\ref{approx3}) and the other term is suppressed by a
power of $\epsilon$. Together with (\ref{U0}), we obtain that
\begin{equation}
\kappa_<\simeq\frac{\Gamma^2\left(\frac{1}{4}\right)}{\pi^{\frac{3}{2}}}
\int_0^{\omega_0}d\omega\ln\rho(\omega)
\end{equation}
with
\begin{equation}
\rho(\omega)=\frac{3^{\frac{3}{2}}C_3^4(\omega)}
{C_1(\omega)C_2^{\frac{1}{2}}(\omega)\sinh^{\frac{5}{2}}2K\omega}
\label{rho}
\end{equation}
and the approximation becomes exact in the limit $\epsilon\to 0$.
It follows from the asymptotic behaviors (\ref{largeomega1}),
(\ref{largeomega2}) and (\ref{largeomega3}) that
\begin{equation}
\ln\rho(\omega)=o\left(\frac{1}{\omega}\right)
\end{equation}
for $\omega>>1$ and
\begin{equation}
\rho(\omega)\simeq\frac{64\sqrt{2}}{\pi\Gamma^2\left(\frac{1}{4}
\right)}\omega\simeq 2.19171\omega
 \label{smallomega}
\end{equation}
as $\omega\to 0$. The very fact that the integration of $\kappa_<$
converges in the limit $\omega_0\to\infty$ indicate that $\kappa_>$
vanishes under the same limit. This is indeed the case. For the
integrand of $\kappa_>$, the approximations
(\ref{approx1})-(\ref{approx3}) cease to be valid because
$\omega\epsilon$ may be of the order one or larger. Treating
$\epsilon$ as a variable and making use of the expansion formula
\begin{equation}
{\rm cn}(-K+\epsilon)={\rm cn}(K-\epsilon)
=\frac{\epsilon}{\sqrt{2}}\left(1-\frac{\epsilon^4}{40}+\frac{\epsilon^8}{1290}+...\right)
\label{quartic1}
\end{equation}
and the identity
\begin{equation}
{\rm sn}^2\sigma{\rm dn}^2\sigma=\frac{1}{2}(1-{\rm cn}^4\sigma)
\label{quartic2}
\end{equation}
we find the approximations of $D_1(\omega)$, $D_2(\omega)$ and $D_\pm(\omega)$ in terms of
${\cal D}_0(\omega)$ and ${\cal D}_1(\omega)$ of the single Wilson line, i. e.
\begin{equation}
\left.D_1(\omega)\right |_{\sigma=\pm(\epsilon-K)}\simeq
\left.D_2(\omega)\right |_{\sigma=\pm(\epsilon-K)} \simeq \left.
D_\pm(\omega) \right |_{\sigma= \pm(\epsilon-K)} \simeq  \left.{\cal
D}_1(\omega)\right |_{\zeta=\epsilon}
\end{equation}
and
\begin{equation}
\left.D_\pm(\omega)\right
|_{\sigma=\pm(K-\epsilon)}\simeq\left.{\cal D}_0(\omega)\right
|_{\zeta=\epsilon}.
\end{equation}
The correction is of the order $\epsilon^4$ which remains small
throughout the integration domain of $\kappa_>$. The WKB analysis
of the appendix A yields
\begin{equation}
\eta_{1,2}(K-\epsilon)=\xi_{1,2}(-K+\epsilon)=C_{1,2}(\omega)v_1(\omega\epsilon)
\Big[1+o\left(\frac{1}{\omega}\right)\Big],
\label{asympt12}
\end{equation}
\begin{equation}
\eta_+(K-\epsilon)=\xi_-(-K+\epsilon)=3C_3(\omega)v_0(\omega\epsilon)
\Big[1+o\left(\frac{1}{\omega}\right)\Big]
\label{asympt3}
\end{equation}
and
\begin{equation}
\eta_-(K-\epsilon)=\xi_+(-K+\epsilon)=C_3(\omega)v_1(\omega\epsilon)
\Big[1+o\left(\frac{1}{\omega}\right)\Big].
\label{asympt4}
\end{equation}
with $C$'s given by the first two terms of their asymptotic
expansions (\ref{largeomega1})-(\ref{largeomega3}). Substituting
eqs.(\ref{asympt12})-(\ref{asympt4}) into the expression of
$U_\alpha(\omega)$, we observe that only one term of the numerator
of (\ref{lambda}) dominates exponentially. It follows from
(\ref{finalR1}) and (\ref{R2}) that
\begin{equation}
{\cal R}_2(\omega)={\cal R}_1^2(\omega)\Big[1+o\left(\frac{1}{\omega}\right)\Big]
\end{equation}
where we have utilized the relations in (\ref{CtoW}) and the
explicit forms of the functions $v$'s in (\ref{uv1})-(\ref{uv-}).
Consequently $\lim_{\omega_0\to\infty}\kappa_>=0$ and
we arrive at the integral representation of the coefficient
$\kappa_1$,
\begin{equation}
\kappa_1=\frac{\Gamma^2\left(\frac{1}{4}\right)}{\pi^{\frac{3}{2}}}
\int_0^{\infty}d\omega\ln\rho(\omega).
\label{finalkappa}
\end{equation}
with $\rho(\omega)$ given by (\ref{rho}). This integral is well
defined and will be evaluated numerically in the next section.

\section{The evaluation of $\kappa_1$ ---- numerical part}

As was explained in section II, the algebraic coordinate $z$ will
introduce an artificial singularity to the differential equations
underlying the determinant ratio of the parallel lines.
The conformal metric (\ref{conformal}) we work with involves elliptic
functions. This is not a big deal for numerical analysis. The elliptic
functions can be expressed as the ratios of theta functions \cite{Whittaker}\cite{Wang},
\begin{equation}
{\rm sn}\sigma=\frac{{\vartheta}_3{\vartheta}_1\left(\frac{\sigma}{2K}\right)}
{{\vartheta}_2{\vartheta}_4\left(\frac{\sigma}{2K}\right)},
\end{equation}
\begin{equation}
{\rm cn}\sigma=\frac{{\vartheta}_4{\vartheta}_2\left(\frac{\sigma}{2K}\right)}
{{\vartheta}_2{\vartheta}_4\left(\frac{\sigma}{2K}\right)},
\end{equation}
and
\begin{equation}
{\rm dn}\sigma=\frac{{\vartheta}_4{\vartheta}_3\left(\frac{\sigma}{2K}\right)}
{{\vartheta}_3{\vartheta}_4\left(\frac{\sigma}{2K}\right)},
\end{equation}
where
\begin{equation}
{\vartheta}_1(z)=2\sum_{n=0}^\infty(-)^n q^{\left(n+\frac{1}{2}\right)^2}
\sin(2n+1)\pi z
\label{theta1}
\end{equation}
\begin{equation}
{\vartheta}_2(z)=2\sum_{n=0}^\infty q^{\left(n+\frac{1}{2}\right)^2}
\cos(2n+1)\pi z,
\label{theta2}
\end{equation}
\begin{equation}
{\vartheta}_3(z)=1+2\sum_{n=1}^\infty q^{n^2}\cos 2n\pi z,
\label{theta3}
\end{equation}
and
\begin{equation}
{\vartheta}_4(z)=1+2\sum_{n=1}^\infty(-)^n q^{n^2}\cos 2n\pi z,
\label{theta4}
\end{equation}
with ${\vartheta}_i\equiv\vartheta_i(0)$ for $i=1,2,3,4$. The
quantity $\vartheta_3$ is related to the complete elliptic
integral of the first kind via $K(k)=\frac{\pi}{2}\vartheta_3^2$
with $k$ the modulo. The expansion parameter $q=e^{-\pi}\simeq
0.0432139$ for our case, $k=\frac{1}{\sqrt{2}}$, so the series
(\ref{theta1})-(\ref{theta4}) converge extremely fast. Also in
this case, $\vartheta_2=\vartheta_4=2^{-\frac{1}{4}}\vartheta_3$.
The Schroedinger like equations
(\ref{potential1})-(\ref{potential+}) are solved with the fourth
order Runge-Kutta method under the boundary conditions
(\ref{etadef}) -(\ref{cond2}). For eqs.(\ref{potential1}) and
(\ref{potential2}), we take advantage of the symmetry property
(\ref{symmetry1}) and evaluate the Wronskian by the formula
\begin{equation}
W_{1,2}(\omega)=-2\eta_{1,2}(0)\eta_{1,2}^\prime(0)
\label{formulaW}
\end{equation}
where the prime denotes the derivative with respect to $\sigma$.
The coefficients $C_{1,2}(\omega)$ follows from (\ref{CtoW}). For
eq.(\ref{potential+}) with the upper sign, we develop
$\eta_+(\sigma)$ from $\sigma\simeq -K$ and $\xi_+(\sigma)$ from
$\sigma=K$, evaluate their Wronskian at $\sigma=0$ and calculate
the coefficient $C_3(\omega)$ from eq.(\ref{CtoW}). An alternative
way is to run the solution $\eta_+(\sigma)$ all the way to $K$ and
calculate the Wronskian by eq.(\ref{alternative}). To avoid the
rapid changes of the potential function near the singularity
$\sigma=-K$, we start with an analytical approximation of
$\eta_{1,2}(\sigma)$ and $\eta_+(\sigma)$ at $\sigma=-K+\delta$
with $\delta<<1$ and then run the Runge-Kutta iteration for
$\sigma>-K+\delta$. Notice that $\delta$ here is not the
regularization parameter introduced below (\ref{s1}) and on LHS of
(\ref{regparrelation})).On writing $x=\omega\delta$, we find the
approximate solutions
\begin{equation}
\eta_{1,2}(-K+\delta)=3\lbrace u_1(x)+c_{1,2}[p(x)u_1(x)+q(x)v_1(x)]\rbrace
\label{near1}
\end{equation}
and
\begin{equation}
\eta_+(-K+\delta)=3\lbrace u_1(x)+c_+[p(x)u_1(x)+q(x)v_1(x)]\rbrace
\label{near2}
\end{equation}
where $u_1$ and $v_1$ are given by (\ref{uv1}),
\begin{equation}
p(x)=\frac{1}{20\omega^4}\Big[\frac{1}{3}x^3-x+\frac{5}{4}
-\left(\frac{1}{2}x^2+\frac{3}{2}x+\frac{5}{4}\right)e^{-2x}\Big]
\label{near3}
\end{equation}
and
\begin{equation}
q(x)=-\frac{1}{20\omega^4}\Big[\frac{1}{3}x^3-x+\frac{1}{2}\left(x^2
+\frac{5}{2}\right)\sinh 2x-\frac{3}{2}x\cosh 2x\Big].
\label{near4}
\end{equation}
The coefficients $c_1=1$, $c_2=-4$ and
$c_+=-\frac{1}{4}$. No such a precaution is necessary for the
solution $\xi_+(\sigma)$ and the Runge-Kutta can start right at
the point $\sigma=K$.The numerical results of $C_1(\omega)$,
$C_2(\omega)$ and $C_3(\omega)$ are displayed in Fig.1, where we have
introduced
\begin{equation}
\hat
C_1(\omega)\equiv\frac{2}{3}C_1(\omega)e^{-2K\omega}=1-\frac{c}{\omega}+...,
\label{large1}
\end{equation}
\begin{equation}
\hat
C_2(\omega)\equiv\frac{2}{3}C_2(\omega)e^{-2K\omega}=1-\frac{2c}{\omega}+...,
\label{large2}
\end{equation}
and
\begin{equation}
\hat C_3(\omega)\equiv
2C_3(\omega)e^{-2K\omega}=1-\frac{c}{2\omega}+... \label{large3}
\end{equation}
with the last step of each equation following from the asymptotic
expansions (\ref{largeomega1}), (\ref{largeomega2}) and
(\ref{largeomega3}). The comparison of the numerical results with
the asymptotic expansions (\ref{large1}), (\ref{large2}) and
(\ref{large3}) is shown in the table 1 and that with the small
$\omega$ behaviors (\ref{barc1}), (\ref{barc2}) and (\ref{barc3})
is shown in the table 2. The agreement is excellent. To gain more
confidence on the numerical solutions of the differential
equations (\ref{potential1}), (\ref{potential2}) and
(\ref{potential+}) for intermediate $\omega$, we checked the
numerical code against a soluble model in which we base the
covariant derivatives in (\ref{parallel}) on the following AdS$_2$
metric
\begin{equation}
ds^2=\frac{1}{\cos^2\sigma}(d\tau^2+d\sigma^2)
\label{ads2}
\end{equation}
with $-\frac{\pi}{2}\le\sigma\le\frac{\pi}{2}$. The differential
equations corresponding to (\ref{potential1}), (\ref{potential2})
and (\ref{potential+}) can be reduced to hypergeometric equations
and the exact forms of the $C$-coefficients of soluble model are
derived in the appendix C. They read
\begin{equation}
C_1^{\rm sol.}(\omega)=C_2^{\rm sol.}(\omega)=\frac{3\omega^2}{\omega^2+1}\sinh\pi\omega
\label{soluble12}
\end{equation}
and
\begin{equation}
C_3^{\rm sol.}(\omega)=\frac{4\omega^2}{4\omega^2+1}\cosh\pi\omega.
\label{soluble3}
\end{equation}
We have
\begin{equation}
\ln\rho^{\rm sol.}(\omega)\equiv\ln\frac{3^{\frac{3}{2}}C_3^{\rm
sol.}(\omega)^4} {C_1^{\rm sol.}(\omega)C_2^{\rm
sol.}(\omega)^{\frac{1}{2}}\sinh^{\frac{5}{2}}\pi\omega}
=\frac{1}{2\omega^2}+O(\frac{1}{\omega^4}) \label{solrho}
\end{equation}
and
\begin{equation}
\int_0^\infty d\omega\ln\rho^{\rm sol.}(\omega)=0. \label{solres}
\end{equation}

In Fig.2, we plot the function $\rho(\omega)$ of
(\ref{rho}) along with that of the soluble model $\rho^{\rm
sol.}(\omega)$. The small $\omega$ behavior of the former can be
fitted to a polynomial
\begin{equation}
\rho(\omega)=2.19171\omega-3.4445\omega^3+6.21735\omega^5-10.8863\omega^7
+17.5978\omega^9,
\end{equation}
consistent with (\ref{smallomega}). For large $\omega$, the
products $\omega^2\rho(\omega)$ and $\omega^2\rho^{\rm
sol.}(\omega)$ are tabulated in table 3 with both determined
numerically. Both Fig.2 and Table 3 suggest that $\rho(\omega)$
falls off faster than $\rho^{\rm sol.}(\omega)$ for large
$\omega$. An analytical demonstration requires extending the WKB
approximation in appendix A to higher orders and will be rather
tedious. Here we merely post our observation without offering a
rigorous proof. The self-adaptive Simpson integration of
$\ln\rho(\omega)$ yields
\begin{equation}
\int_0^\infty d\omega\ln\rho(\omega)\simeq -0.56534,
\label{realres}
\end{equation}
which, upon substitution into (\ref{finalkappa}) leads to
\begin{equation}
\kappa_1=-1.33460. \label{kappa1}
\end{equation}

The relative error in the numerical valuation of the elliptic
functions is about $10^{-15}$ and that of the coefficients $C^{\rm
sol.}(\omega)$'s extracted from our Runge-Kutta iteration is found
below $10^{-11}$. Notice that the near singularity expansion of
the trigonometric functions pertaining to the soluble model goes
by the second power of $\epsilon$, while the same type of
expansion of the elliptic functions pertaining to the parallel
lines, (\ref{quartic1}) and (\ref{quartic2}), goes by the fourth
power of $\epsilon$. Therefore the approximations
(\ref{near1})-(\ref{near4}) should work better for the parallel
lines. Likewise is the numerical integration (\ref{realres}), the
integrand of which vanishes faster than that of the soluble model
at large $\omega$. For the soluble model, we found $2.39\times
10^{-8}$ in contrast to the exact value zero of (\ref{solres}).
Consequently, the accuracy of our numerical algorithm should be
amply sufficient for the six significant figures of the $\kappa$
value reported in this paper.

\begin{center}
\begin{table}
\begin{tabular}{|c|c|c|c|c|c|c|c|c|c|c|}\hline
$\omega$&10&100&200&300&400&500&600&700\\
\hline $\omega(1-\hat C_1(\omega))$&0.848635&0.847228&0.847217&0.847215&0.847214&0.847214&0.847214&0.847213\\
\hline $\frac{\omega}{2}(1-\hat C_2(\omega))$
&0.844365&0.847182&0.847205&0.847210&0.847211&0.847212&0.847212&0.847212\\
\hline $2\omega(1-\hat C_3(\omega))$
&0.846503&0.847205&0.847211&0.847212&0.847213&0.847213&0.847213&0.847213\\\hline
\end{tabular}
\caption{The large $\omega$ behaviors of the numerically generated
$\hat C_1(\omega)$, $\hat C_2(\omega)$ and $\hat C_3(\omega)$}
\end{table}
\end{center}

\begin{center}\begin{table}
\begin{tabular}{|c|c|c|c|c|c|c|c|c|c|}\hline
$\omega$&0.00001&0.00005&0.0001&0.0005&0.001&0.005&0.01\\
\hline
$\frac{C_1(\omega)}{\omega^3}$&10.16655701&10.16655704&10.16655711&10.16655950&10.16656698&10.16680621&10.16755383
\\
\hline
$\frac{C_2(\omega)}{\omega^3}$&5.08327851&5.08327852&5.08327857&5.08328004&5.08328465&5.08343202&5.08389259\\\hline
$\frac{C_3(\omega)}{\omega^2}$&4.00000000&4.00000001&4.00000006&4.00000143&4.00000574&4.00014355&4.00057424\\
\hline
\end{tabular}
\caption{The small $\omega$ behaviors of the numerically generated $C_1(\omega)$, $C_1(\omega)$ and $C_1(\omega)$}
\end{table}\end{center}

\begin{center}\begin{table}
\begin{tabular}{|c|c|c|c|c|c|c|c|c|c|c|}\hline
$\omega$&10&100&200&300&400&500&600&700\\
\hline $\omega^2\ln {\rho(\omega)}$&-0.000052&-0.000001&-0.000002&-0.000003&-0.000004&-0.000004&-0.000005&-0.000005\\
\hline $\omega^2\ln
{\rho^{sol.}(\omega)}$&0.493798&0.499938&0.499984&0.499993&0.499996&0.499997&0.499998&0.499999\\\hline
\end{tabular}
\caption{The large $\omega$ behaviors of $\rho(\omega)$ and
$\rho^{\rm sol.}(\omega)$.}
\end{table}\end{center}

\begin{figure}
\includegraphics[width=0.50\textwidth]{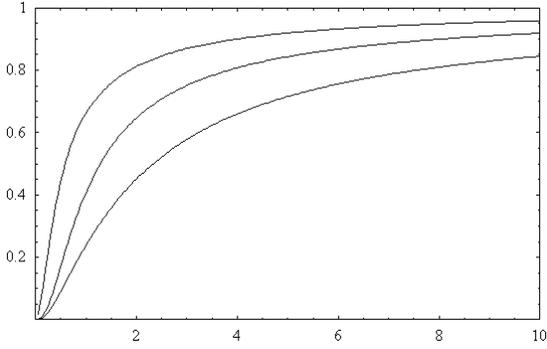}
\caption{the top curve represents $\hat C_3(\omega)$, the middle one
represents $\hat C_1(\omega)$, the bottom one represents $\hat
C_2(\omega)$. }
\end{figure}

\begin{figure}
  \centering
    \includegraphics[width=0.49\textwidth]{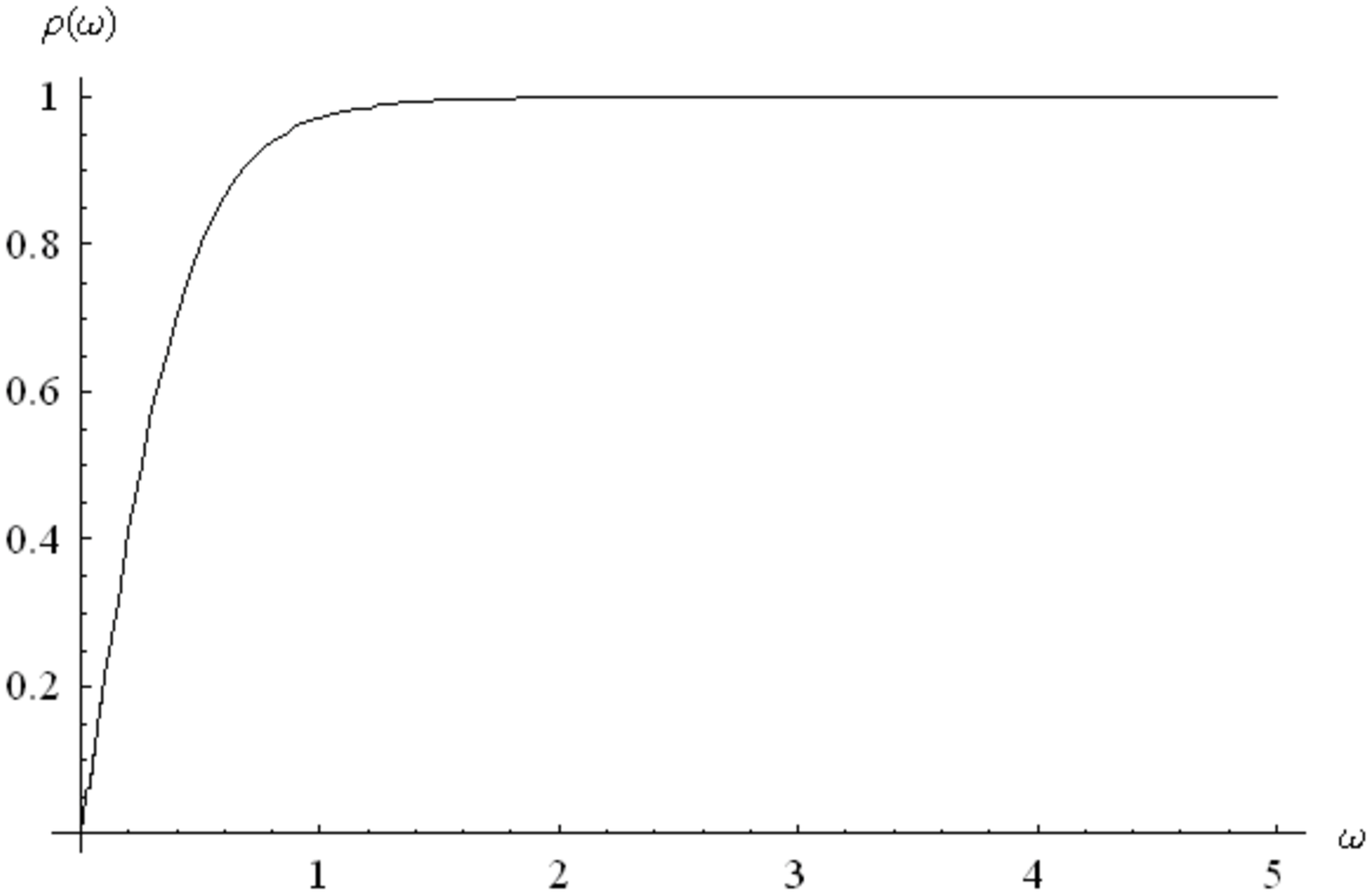}
    \includegraphics[width=0.49\textwidth]{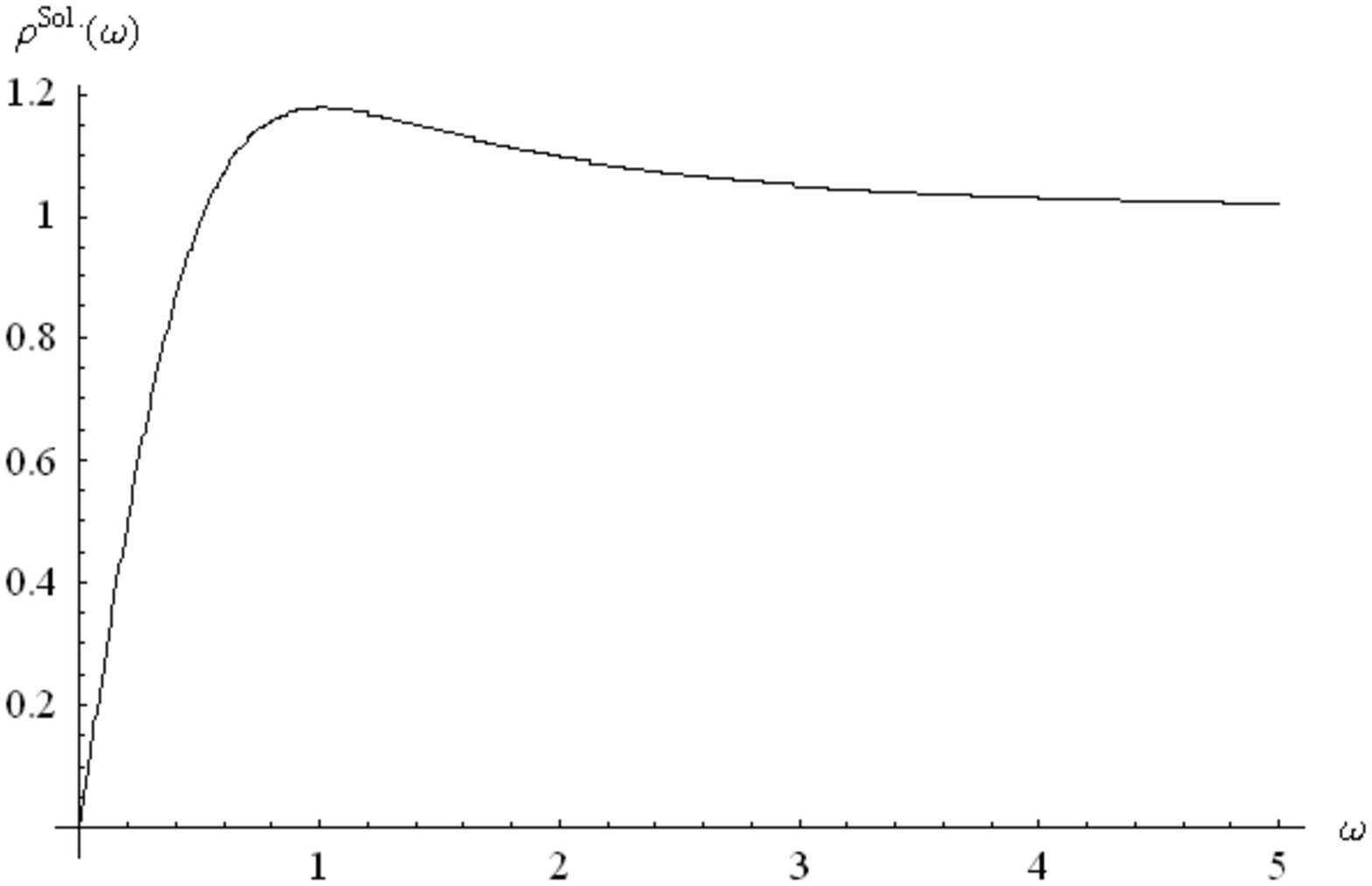}
  \caption{the left curve represents $\rho(\omega)$,
while the right one represents $\rho^{\rm{sol.}}(\omega)$.}
\end{figure}


\section{Determination of $\kappa_2$}


To determine $\kappa_2$, we quote two formula of Ref.\cite{Drukker:2000ep}, one for a bosonic
determinant and the other for a fermionic determinant. Consider a general 2d metric
\begin{equation}
ds^2=g_{\alpha\beta}d\sigma^\alpha d\sigma^\beta
\label{2dmetric}
\end{equation}
with the scalar curvature $R$. Define the functional operator
$\Delta_M\equiv M^{-1}(-\nabla^2+X)$ with $\nabla^2$ the Laplacian
with respect to the metric (\ref{2dmetric}) and ($M$, $X$)
functions of coordinates. Varying $M$ amounts to a conformal
transformation of the metric (\ref{2dmetric}) and associated anomaly
contributes a nontrivial finite term to the variation
of the functional determinant of $\Delta_M$. We have
\begin{equation}
\left(\ln\frac{\det\Delta_M)}{\det\Delta_1}\right)_{\rm fin.}
=-\frac{1}{4\pi}\int d^2\sigma\sqrt{g}\Big[\ln M\left(\frac{1}{6}R-X\right)
+\frac{1}{12}g^{\alpha\beta}\frac{\partial\ln M}{\partial\sigma^\alpha}
\frac{\partial\ln M}{\partial\sigma^\beta}\Big]+\hbox{boundary terms},
\label{drukker1}
\end{equation}
Since we are always taking the difference between the parallel Wilson lines and the two
single Wilson lines, the boundary terms cancel and we may integrate by part freely.
For a Dirac operator with respect to the metric (\ref{2dmetric}),
$\gamma^\alpha\nabla_\alpha$, we define
\begin{equation}
\Delta_{\cal K}^F\equiv-({\cal
K}^{-1}\gamma^\alpha\nabla_\alpha)^2+{\cal K}^{-2}Y
\end{equation}
with ${\cal K}$ and $Y$ functions of coordinates. The measure
transformation formula corresponding to (\ref{drukker1}) reads.
\begin{equation}
\left(\ln\frac{\det\Delta_{\cal K}^F)}{\det\Delta_1^F}\right)_{\rm fin.}
=\frac{1}{2\pi}\int d^2\sigma\sqrt{g}\Big[\ln{\cal K}\left(\frac{1}{6}R+2Y\right)
+\frac{1}{6}g^{\alpha\beta}\frac{\partial\ln{\cal K}}{\partial\sigma^\alpha}
\frac{\partial\ln{\cal K}}{\partial\sigma^\beta}\Big]+\hbox{boundary terms},
\label{drukker2}
\end{equation}
where we have multiplied the integral in \cite{Drukker:2000ep} by
two, taking into account that $\Delta_{\cal K}^F$ here is a $2\times
2$ matrix in the spinor space.

Coming to the determinants we are interested in, metric
(\ref{conformal1})and metric (\ref{conformal}) are all conformal
with
\begin{equation}
g_{\alpha\beta}=e^{-2\chi}\delta_{\alpha\beta}
\end{equation}
and the scalar curvature
\begin{equation}
R=2e^{2\chi}\delta^{\alpha\beta}\frac{\partial^2\chi}{\partial\sigma^\alpha\partial\sigma^\beta}.
\end{equation}
We have $(\sigma_0,\sigma_1)=(\tau,\zeta)$ and $\chi=\ln\zeta$ for
the single line, and $(\sigma_0,\sigma_1)=(\tau,\sigma)$ and
$\chi=\ln(\sqrt{2}{\rm cn}\sigma)$ for the parallel lines. With
the measure scaling functions $M=e^{2\chi}$ and ${\cal K}=e^\chi$,
those functional operators of (\ref{Delta01})-(\ref{DF2}) without
hats corresponds to $\Delta_1$ and $\Delta_1^F$ of
eqs.(\ref{drukker1}) and (\ref{drukker2}) and that with hats to
$\Delta_M$ and $\Delta_{\cal K}^F$ there. The "mass square" $X$ of
(\ref{drukker1}) equals to zero for $\Delta_0[{\cal C}_1]$ and
$\Delta_0[{\cal C}_2]$, equals to 2 for $\Delta_1[{\cal C}_1]$ and
$\Delta_1[{\cal C}_2]$ and equals to $4+R$ for $\Delta_2[{\cal
C}_2]$. The "mass" $Y$ of (\ref{drukker2}) equals to one for all
fermionic determinants. It follows from (\ref{drukker1}) that
\begin{equation}
\left(\ln\frac{\det\Delta_0[{\cal C}_1]}{\det\hat\Delta_0[{\cal C}_1]}\right)_{\rm fin.}
=-\frac{1}{12\pi}\int_{{\cal C}_1}d^2\sigma\delta^{\alpha\beta}
\frac{\partial\chi}{\partial\sigma^\alpha}\frac{\partial\chi}{\partial\sigma^\beta}
+\hbox{boundary terms},
\end{equation}
\begin{equation}
\left(\ln\frac{\det\Delta_0[{\cal C}_2]}{\det\hat\Delta_0[{\cal C}_2]}\right)_{\rm fin.}
=-\frac{1}{12\pi}\int_{{\cal C}_2}d^2\sigma\delta^{\alpha\beta}
\frac{\partial\chi}{\partial\sigma^\alpha}\frac{\partial\chi}{\partial\sigma^\beta}
+\hbox{boundary terms},
\end{equation}
\begin{equation}
\left(\ln\frac{\det\Delta_1[{\cal C}_1]}{\det\hat\Delta_1[{\cal C}_1]}\right)_{\rm fin.}
=-\frac{1}{12\pi}\int_{{\cal C}_1}d^2\sigma\delta^{\alpha\beta}
\frac{\partial\chi}{\partial\sigma^\alpha}\frac{\partial\chi}{\partial\sigma^\beta}
-\frac{1}{\pi}\int_{{\cal C}_1}d^2\sigma e^{-2\chi}\chi
+\hbox{boundary terms},
\end{equation}
\begin{equation}
\left(\ln\frac{\det\Delta_1[{\cal C}_2]}{\det\hat\Delta_1[{\cal C}_2]}\right)_{\rm fin.}
=-\frac{1}{12\pi}\int_{{\cal C}_2}d^2\sigma\delta^{\alpha\beta}
\frac{\partial\chi}{\partial\sigma^\alpha}\frac{\partial\chi}{\partial\sigma^\beta}
-\frac{1}{\pi}\int_{{\cal C}_2}d^2\sigma e^{-2\chi}\chi
+\hbox{boundary terms},
\end{equation}
and
\begin{equation}
\left(\ln\frac{\det\Delta_2[{\cal C}_2]}{\det\hat\Delta_2[{\cal C}_2]}\right)_{\rm fin.}
=\frac{11}{12\pi}\int_{{\cal C}_2}d^2\sigma\delta^{\alpha\beta}
\frac{\partial\chi}{\partial\sigma^\alpha}\frac{\partial\chi}{\partial\sigma^\beta}
-\frac{2}{\pi}\int_{{\cal C}_2}d^2\sigma e^{-2\chi}\chi
+\hbox{boundary terms}.
\end{equation}
where the subscript of the integration sign indicates the world sheet integration extends to.
Similarly, the formula (\ref{drukker2}) implies that
\begin{equation}
\left(\ln\frac{|\det D_F[{\cal C}_1]|}{|\det\hat D_F[{\cal C}_1]|}\right)_{\rm fin.}
=\frac{1}{24\pi}\int_{{\cal C}_1}d^2\sigma\delta^{\alpha\beta}
\frac{\partial\chi}{\partial\sigma^\alpha}\frac{\partial\chi}{\partial\sigma^\beta}
-\frac{1}{2\pi}\int_{{\cal C}_1}d^2\sigma e^{-2\chi}\chi
+\hbox{boundary terms}
\end{equation}
and
\begin{equation}
\left(\ln\frac{|\det D_F[{\cal C}_2]|}{|\det\hat D_F[{\cal C}_2]|}\right)_{\rm fin.}
=\frac{1}{24\pi}\int_{{\cal C}_2}d^2\sigma\delta^{\alpha\beta}
\frac{\partial\chi}{\partial\sigma^\alpha}\frac{\partial\chi}{\partial\sigma^\beta}
-\frac{1}{2\pi}\int_{{\cal C}_2}d^2\sigma e^{-2\chi}\chi
+\hbox{boundary terms}.
\end{equation}
Substituting into (\ref{DeltaS21}) and (\ref{DeltaS22}) for the
single line and the parallel lines, we find their contributions
add up to zero in each case i.e. $W_2[{\cal C}_1]=W_2[{\cal
C}_2]=0$. Consequently,
\begin{equation}
\kappa_2=0.
\label{kappa2}
\end{equation}
This, together with (\ref{kappa1}) leads to our final result
(\ref{finalresult}).


\section{Concluding remarks}


As AdS/CFT has become an important reference to understand the
observation of the strongly interacting quark-gluon plasma created
by heavy ion collisions, it is critical to asses the robustness of
the leading order prediction by exploring the next order correction
in the expansion according to the inverse powers of the large 't
Hooft coupling $\lambda=N_cg_{\rm YM}^2$. The subleading terms of
the expansion have been addressed in the literature in the context
of the equation of state \cite{Gubser:1998nz} and the shear
viscosity \cite{Buchel:2004di}\cite{Buchel:2008sh}. This type of
corrections comes from the $\alpha^{\prime 3}$ correction of the
target space metric \cite{Pawelczyk:1998pb}. Its contribution is of
the order $O(\lambda^{-3/2})$ relative to the leading order in the
${\cal N}=4$ SYM and is present only at nonzero temperature. In case
of the expectation value of a Wilson loop operator, however, the
dominant correction stems from the fluctuation of the world sheet
around its minimum area and is suppressed only by
$O(\lambda^{-1/2})$ relative to the leading order. It shows up at
all temperatures and is more difficult to compute. The only attempts
made in the literature in this regard include the strong coupling
expansion of a single line, a circular loop and a spinning line at
zero temperature
\cite{Forste:1999qn}\cite{Drukker:2000ep}\cite{Kruczenski:2008zk}\cite{frolov}.
These Wilson loops, though theoretically important, do not carry
direct phenomenological implications.

In this work, we have extended the method in \cite{Kruczenski:2008zk} to the fluctuations of the world sheet
dual to a pair of parallel Wilson lines and have derived the next term of the strong coupling expansion of the
heavy quark-antiquark potential in ${\cal N}=4$ SYM at zero temperature.
We start with the determinant ratio for a single Wilson line and that for parallel lines in the static gauge, in which
the fluctuations come from eight transverse bosonic coordinates and eight 2d Majorana fermions.
Then we scaled the operators underlying the determinants, leaving a trivial measure for
the associated spectral problem. The subleading term of the heavy quark potential is extracted
from the combination (\ref{oneloop}), which consists of the spectral and the measure parts. A robust
numerical result of the former is obtained and the contributions from measure change of each determinant
cancel. We have,
\begin{equation}
V(r)=-\frac{a(\lambda)}{r}
\end{equation}
with
\begin{equation}
a(\lambda)=\left\{\begin{array}{ll}
\begin{gathered}
\frac{4\pi^2}{\Gamma^4\left(\frac{1}{4}\right)}\frac{\sqrt{\lambda}}{r}
\Big[1-\frac{1.33460}{\sqrt{\lambda}}+O\left(\frac{1}{\lambda}\right)\Big],
\hspace{0.2cm}\hbox{for $\lambda>>1$}\\
\end{gathered}
\hfill\\
\begin{gathered}
\frac{\lambda}{4\pi r}\Big[1-\frac{\lambda}{2\pi^2}
\left(\ln\frac{2\pi}{\lambda}-\gamma_E+1\right)+O(\lambda^2)\Big],
\hspace{0.2cm}\hbox{for $\lambda<<1$}\\
\end{gathered}
\end{array}
\right .  \label{expansion1}
\end{equation}
where the weak coupling expansion obtained in \cite{erickson}\cite{pineda} from field theory is also
included for completeness.
The authors of \cite{erickson} also worked out the strong coupling expansion under
the ladder approximation in field theory,
\begin{equation}
V_{\rm ladder}(r)=-\frac{\sqrt{\lambda}}{\pi
r}\left(1-\frac{\pi}{\sqrt{\lambda}}\right).
\end{equation}
It is interesting to notice that our subleading term is of the same sign as theirs but the magnitude
relative to the leading order is smaller in our result.
In view of the range of the 't Hooft coupling which was used for the RHIC phenomenology,
\begin{equation}
5.5<\lambda<6\pi
\label{range}
\end{equation}
the correction to the leading order of the strong coupling may be significant in magnitude.
One may define an effective coupling
\begin{equation}
\sqrt{\lambda^\prime}=\sqrt{\lambda}-1.33460
\end{equation}
If $\lambda$ of (\ref{range}) is replaced by $\lambda^\prime$, the range of the 't Hooft
coupling is shifted to
\begin{equation}
13.54<\lambda<32.22
\end{equation}
At a nonzero temperature $T$, however, the order
$O(\lambda^{-1/2})$ is not merely a redefinition of the coupling
and the strong coupling expansion of the heavy quark potential
becomes
\begin{equation}
V(r)\simeq-\frac{4\pi^2}{\Gamma^4\left(\frac{1}{4}\right)}\frac{\sqrt{\lambda}}{r}
\Big[g_0(rT)-\frac{1.33460g_1(rT)}{\sqrt{\lambda}}+O\left(\frac{1}{\lambda}\right)\Big]
\label{expansion1}
\end{equation}
with $g_0(x)$ and $g_1(x)$ two functions satisfying the conditions $g_0(0)=g_1(0)=1$. The function
$g_0(x)$ have been determined by the minimum area of the world sheet in the
Schwarzschild-AdS$_5\times$S$^5$ target space \cite{Rey:1998bq}
\begin{equation}
ds^2=\frac{1}{z^2}\left(f(z)dt^2+d\vec x^2+\frac{1}{f(z)}dz^2\right)+d\Omega_5^2 \label{target}
\end{equation}
with $f(z)=1-\pi^4T^4z^4$ and $t$ the Euclidean time. The one loop effective action underlying
the function $g_1(x)$ has been developed in \cite{Hou:2009zk}
and the methodology employed in this work can be readily generalized there.

While simple in practice, the static gauge we worked with suffers a
problem. Though the combination (\ref{oneloop}) gives rise to a
finite result, neither the UV divergence nor the conformal anomaly
of each term on RHS of (\ref{oneloop}) vanishes. A less problematic
gauge is the conformal gauge, in which the world sheet metric is not
set to the induced metric at the beginning. One has to include the
determinant of the longitudinal fluctuations and that of the ghost
and an appropriate measure of the path integral. The contributions
from the transverse bosons and fermions obtained in this paper will
remain there, but other contributions including the measure change
may be subtle to collect. It is important to carry out the parallel
analysis in the conformal gauge to ascertain that our result in this
paper is complete. Another alternative is the canonical quantization
method employed in \cite{frolov}. We hope to report our progress in
this direction in near future.

\section*{Acknowledgments}

We thank James T. Liu for bring our attention to the Ref.\cite{Kruczenski:2008zk},
which motivated the research reported in this paper. We are grateful to
M. Kruczenski and A. Tirziu for several communications about their
work. We are particularly indebted to A. Tseytlin for clarifying some conceptual
issues in their paper \cite{Drukker:2000ep}. Their comments helped us to
correct an error in a previous version. The valuable suggestions from N. Drukker
are also warmly acknowledged.
The research of D. F. H. and H. C. R. is supported in part
by NSFC under grant Nos. 10575043, 10735040. The work of D. F.
H. is also supported in part by Educational Committee of China
under grant NCET-05-0675 and project No. IRT0624.


\appendix

\section {}

To extract the large $\omega$ behavior of the coefficients
$C_1(\omega)$, $C_2(\omega)$ and $C_3(\omega)$, we introduce
$x\equiv\omega(K+\sigma)$ and $y\equiv\omega(K-\sigma)$. For
$\sigma+K<<1$($K-\sigma<<1$), the solutions of the differential
equations can be approximated by that of the equations for the
single line, which extends to $x>>1$($y>>1$) for large $\omega$.
The WKB approximation applies for $x>>1$ and $y>>1$. In case of
eq.(\ref{potential+}) with upper(lower) sign, the WKB solution can be
extended all the way to the point $\sigma=K$($\sigma=-K$) and the
requirement $y>>1$($x>>1$) may be relaxed. We match the single
line solution and the WKB ones in the regions where both
approximations apply.

Consider the equation (\ref{potential1}) first. We start with the approximate solution near $\sigma=-K$
\begin{equation}
\eta_1\simeq 3u_1(x)
\label{start}
\end{equation}
with $\sigma+K<<1$, where the coefficient 3 follows
from the requirement (\ref{etadef}). The asymptotic form for $x>>1$ reads
\begin{equation}
\eta_1\simeq \frac{3}{2}e^x\left(1-\frac{1}{x}\right)
\simeq\frac{3}{2}e^{x-\frac{1}{x}}.
\label{begin}
\end{equation}
The WKB solution to be matched is given by
\begin{equation}
\eta_1\simeq \exp\left({\int^\sigma d\sigma'\sqrt{\omega^2+\frac{1}{{\rm cn}^2\sigma'}}}\right).
\label{wkb1}
\end{equation}
Expanding the square root for large $\omega$ and using the derivative formula
\begin{equation}
\frac{d}{d\sigma}\frac{{\rm sn}\sigma{\rm dn}\sigma}{{\rm cn}\sigma}
=\frac{1}{2}\left(\frac{1}{{\rm cn}^2\sigma}+{\rm cn}^2\sigma\right)
\end{equation}
we find that
\begin{equation}
\eta_1\simeq A\exp\left(\omega\sigma+\frac{1}{2\omega}\int_{-K}^\sigma\frac{d\sigma'}{{\rm cn}^2\sigma'}
\right)\simeq A\left(\omega\sigma+\frac{{\rm sn}\sigma{\rm dn}\sigma}{\omega{\rm cn}\sigma}
-\frac{1}{2\omega}\int_{-K}^\sigma d\sigma'{\rm cn}^2\sigma'\right)
\label{exponent1}
\end{equation}
with $A$ a constant to be determined. In the left matching region where
$x>>1$ and $\sigma+K<<1$, the approximations
\begin{equation}
\frac{{\rm sn}(\sigma){\rm dn}(\sigma)}{{\rm cn}(\sigma)}
\simeq-\frac{1}{K+\sigma}
\label{approximation}
\end{equation}
and $\int_{-K}^{-\sigma} d\sigma'{\rm cn}^2\sigma'\simeq 0$
yield the coefficient $A=\frac{3}{2}e^{K\omega}$. In the right
matching region where $K-\sigma<<1$ and $y>>1$, the WKB solution (\ref{exponent1}) becomes
\begin{equation}
\eta_1\simeq\frac{3}{2}e^{2K\omega-\frac{c}{\omega}}e^{-y+\frac{1}{y}},
\label{match1}
\end{equation}
where we have used the approximation
\begin{equation}
\frac{{\rm sn}(\sigma){\rm dn}(\sigma)}{{\rm cn}(\sigma)}
\simeq\frac{1}{K-\sigma}
\end{equation}
there and the constant
\begin{equation}
c=\frac{1}{2}\int_{-K}^K d\sigma{\rm cn}^2\sigma
=\sqrt{2}\int_0^1dx\frac{x^2}{\sqrt{1-x^4}}
=\frac{2\pi^{\frac{3}{2}}}{\Gamma^2\left(\frac{1}{4}\right)}.
\end{equation}
Comparing with the expression of (\ref{uv1}), we obtain that
\begin{equation}
\eta_1\simeq\frac{3}{2}e^{2K\omega-\frac{c}{\omega}}v_1(y).
\end{equation}
The asymptotic behavior of (\ref{largeomega1}) is extracted in the limit $y\to 0$
and the relation (\ref{asympt12}) for $\eta_1$ and $\xi_1$
follows.

The equation (\ref{potential2}) can be treated similarly. We start with the same
expression of (\ref{start}) for $\eta_2(\sigma)$ near $\sigma=-K$ but replace the WKB
solution (\ref{wkb1}) by
\begin{equation}
\eta_2\simeq \exp\left(\int^\sigma d\sigma'\sqrt{\omega^2+
\frac{1}{{\rm cn}^2\sigma '} -{\rm cn}^2\sigma '} \right)
\label{wkb2}
\end{equation}
Eqs.(\ref{exponent1}) and (\ref{match1}) become
\begin{equation}
\eta_2\simeq A\exp\left(\omega\sigma+\frac{1}{2\omega}\int_{-K}^\sigma\frac{d\sigma'}{{\rm cn}^2\sigma'}
-\frac{1}{2\omega}\int_{-K}^\sigma d\sigma'{\rm cn}^2\sigma'\right)
\simeq A\exp\left(\omega\sigma+\frac{{\rm sn}\sigma{\rm dn}\sigma}{{\rm cn}\sigma}
-\frac{1}{\omega}\int_{-K}^\sigma d\sigma'{\rm cn}^2\sigma'\right)
\label{exponent2}
\end{equation}
with the same $A$ and
\begin{equation}
\eta_2(\sigma)\simeq\frac{3}{2}e^{2K\omega-\frac{2c}{\omega}}e^{-y+\frac{1}{y}}
\label{match2}
\end{equation}
for $y>>1$ and $K-\sigma<<1$. The asymptotic behavior
(\ref{largeomega2}) and the relation (\ref{asympt12}) for
$(\eta_2,\xi_2)$ are extracted then.

Coming to eq.(\ref{potential+}), the single line solution
(\ref{start}) remains approximating and we have
\begin{equation}
\eta_+\simeq 3u_1(x)
\end{equation}
for $\sigma+K<<1$. The WKB solution it matches with for $x>>1$ reads
\begin{eqnarray}
\eta_+ &\simeq& \exp\left({\int^\sigma d\sigma'\sqrt{\omega^2+\frac{1}{2{\rm cn}^2\sigma'}
-\frac{{\rm sn}\sigma'{\rm dn}\sigma'}{\sqrt{2}{\rm
cn}^2\sigma'}}}\right) \nonumber\\
&\simeq&\frac{3}{2}
\exp\left(K\omega+{\omega\sigma+\frac{1}{4\omega}\int^\sigma d\sigma'
\frac{1-\sqrt{2}{\rm sn}\sigma'{\rm dn}\sigma'}{{\rm cn}^2\sigma'}}\right)\nonumber\\
&\simeq&\frac{3}{2}
\exp\left(K\omega+{\omega\sigma+\frac{1}{2\sqrt{2}\omega}
\frac{\sqrt{2}{\rm sn}\sigma{\rm dn}\sigma-1}{{\rm cn}\sigma}
-\frac{1}{4\omega}\int_{-K}^\sigma d\sigma'{\rm cn}^2\sigma'}\right)
\label{wkb3}
\end{eqnarray}
and works all the way to the point $\sigma=K$. Near that point, we find
\begin{equation}
\eta_+\simeq \frac{3}{2}e^{2K\omega-\frac{c}{2\omega}-y}
\end{equation}
Eqs.(\ref{largeomega3}) and (\ref{asympt3}) follow from the form of $v_0(y)$, (\ref{CtoW}) and
(\ref{alternative}). The relation (\ref{asympt4}) is obtained starting with the WKB solution
\begin{equation}
\xi_+\simeq \sinh\left(\int_\sigma^K
d\sigma'\sqrt{\omega^2+\frac{1}{2{\rm cn}^2\sigma'} -\frac{{\rm
sn}\sigma'{\rm dn}\sigma'}{\sqrt{2}{\rm cn}^2\sigma'}}\right)
\end{equation}
and matching it to the approximate solution $v_1(x)$ for
$K+\sigma<<1$ and $x>>1$.

\section {}

The differential equation (\ref{potential1}) at $\omega=0$ can be converted to a
hypergeometric equation by the transformation
\begin{equation}
x={\rm cn}^4\sigma \qquad \phi(\sigma)=\sqrt{x}f(x),
\label{hyper}
\end{equation}
i.e.
\begin{equation}
x(1-x)\frac{d^2f}{dx^2}+\left(\frac{7}{4}-\frac{9}{4}x\right)\frac{df}{dx}-\frac{3}{8}f=0.
\end{equation}
We have
\begin{equation}
\bar\eta_1(\sigma)=2{\rm cn}^2\sigma F\left(\frac{1}{2},\frac{3}{4};\frac{7}{4};{\rm cn}^4\sigma\right)
\end{equation}
and $\bar\xi_1(\sigma)=\bar\eta_1(-\sigma)$. It follows from the formula
\begin{equation}
\frac{d}{dz}F(a,b;c;z)=\frac{ab}{c}F(a+1,b+1;c+1;z),
\label{hypergeo1}
\end{equation}
\begin{equation}
F(a,b;c;1)=\frac{\Gamma(c)\Gamma(c-a-b)}{\Gamma(c-a)\Gamma(c-b)}
\label{hypergeo2}
\end{equation}
for ${\rm Re}(c-a-b)>0$ and
\begin{equation}
F(a,b;c;1-\epsilon)=\frac{\Gamma(c)\Gamma(a+b-c)}{\Gamma(a)\Gamma(b)}
\epsilon^{c-a-b}+...
\label{hypergeo3}
\end{equation}
for ${\rm Re}(c-a-b)<0$ and $\epsilon>0$ that the Wronskian
\begin{eqnarray}
W[\bar\eta_1,\bar\xi_1]&=&-2\lim_{\epsilon\to 0^+}\bar\eta_1(-\epsilon)
\bar\eta_1^\prime(-\epsilon)=-\frac{48}{7\sqrt{2}}
\frac{\Gamma\left(\frac{7}{4}\right)\Gamma\left(\frac{1}{2}\right)}{\Gamma\left(\frac{5}{4}\right)\Gamma(1)}
\frac{\Gamma\left(\frac{11}{4}\right)\Gamma\left(\frac{1}{2}\right)}
{\Gamma\left(\frac{3}{2}\right)\Gamma\left(\frac{7}{4}\right)}\nonumber\\
&=&-\frac{72\pi^{\frac{3}{2}}}{\Gamma^2\left(\frac{1}{4}\right)}
\end{eqnarray}
Divided by -3, we derive (\ref{barc1}).

With the same transformation (\ref{hyper}), eq.(\ref{potential2}) becomes
\begin{equation}
x(1-x)\frac{d^2f}{dx^2}+\left(\frac{7}{4}-\frac{9}{4}x\right)\frac{df}{dx}-\frac{1}{4}f=0.
\end{equation}
We have
\begin{equation}
\bar\eta_2(\sigma)=2{\rm cn}^2\sigma F\left(1,\frac{1}{4};\frac{7}{4};{\rm cn}^4\sigma\right)
\end{equation}
and $\bar\xi_2(\sigma)=\bar\eta_2(-\sigma)$. It follows from (\ref{hypergeo1})-(\ref{hypergeo3}) that
the Wronskian
\begin{eqnarray}
W[\bar\eta_2,\bar\xi_2]&=&-2\lim_{\epsilon\to 0^+}\bar\eta_2(-\epsilon)
\bar\eta_2^\prime(-\epsilon)=-\frac{32}{7\sqrt{2}}
\frac{\Gamma\left(\frac{7}{4}\right)\Gamma\left(\frac{1}{2}\right)}
{\Gamma\left(\frac{3}{4}\right)\Gamma\left(\frac{3}{2}\right)}
\frac{\Gamma\left(\frac{11}{4}\right)\Gamma\left(\frac{1}{2}\right)}
{\Gamma(2)\Gamma\left(\frac{5}{4}\right)}\nonumber\\
&=&-\frac{36\pi^{\frac{3}{2}}}{\Gamma^2\left(\frac{1}{4}\right)}
\end{eqnarray}
and (\ref{barc2}) is obtained as $-W[\bar\eta_2,\bar\xi_2]/3$. Using the series representation of
the hypergeometric function, we find that
\begin{equation}
\bar\eta_1=\frac{3}{2}x^{-\frac{1}{4}}\int_0^x\frac{dx'x'^{-\frac{1}{4}}}{\sqrt{1-x'}}
=\frac{3\sqrt{2}}{{\rm cn}\sigma}\int_{-K}^\sigma d\sigma'{\rm cn}^2\sigma'.
\end{equation}
But we fail to find a similar expression for $\bar\eta_2(\sigma)$.

As to $\bar C_3(0)$, we notice that the solution of the 1st order differential equation
\begin{equation}
\sqrt{2}{\rm cn}\sigma\frac{d\psi}{d\sigma}+\psi=0
\label{1stODE}
\end{equation}
also solves eq.(\ref{potential+}) with the upper sign. The eq.(\ref{1stODE}) can be
solved readily and we obtain
\begin{equation}
\psi(\sigma)=B\sqrt{\frac{\sqrt{2}{\rm dn}\sigma-{\rm sn}\sigma}
{\sqrt{2}{\rm dn}\sigma+{\rm sn}\sigma}} \label{barxi}
\end{equation}
with $B$ a constant, where we have used the
indefinite integral
\begin{equation}
\int d\sigma\frac{1}{{\rm cn}\sigma}=-\frac{1}{\sqrt{2}}
\ln\frac{\sqrt{2}{\rm dn}\sigma-{\rm sn}\sigma}{\sqrt{2}{\rm dn}\sigma+{\rm sn}\sigma}+{\rm const.}
\end{equation}
as can be verified by taking derivatives of both sides. Setting
the constant $B=2$, we find that the function $\psi(\sigma)$
satisfies the boundary condition of $\bar\xi_+(\sigma)$ at
$\sigma=K$ and therefore $\bar\xi_+(\sigma)=\psi(\sigma)$. As
$\sigma\to-K$,
\begin{equation}
\bar\xi_+(\sigma)=\frac{4}{\sigma+K}+...
\end{equation}
and we end up with $\bar C_3(0)=4$.

\section {}

In this appendix, we present the details of the soluble model
which is introduced to check our numerical algorithm. The model is
largely motivated by the work in \cite{Tanni}. We shall use the
same symbols $(\eta,\xi)$ for the solutions of the counterparts of
the differential equations (\ref{potential1}), (\ref{potential2})
and (\ref{potential+}). Because the scalar curvature of the metric
(\ref{ads2}) is $R=-2$, the counterparts of (\ref{potential1}) and
(\ref{potential2}) are the same. Consequently,
$D_1(\omega)=D_2(\omega)$ and $(\eta_1,\xi_1) =(\eta_2,\xi_2)$ in
this case. The counterpart of the eq.(\ref{potential1}) or
(\ref{potential2}) reads,
\begin{equation}
-\frac{d^2\phi}{d\sigma^2}+\left(\omega^2+\frac{2}{\cos^2\sigma}\right)\phi=0
\end{equation}
and has the same set of indexes at the regular points
$\sigma=\pm\frac{\pi}{2}$ as that of (\ref{potential1}). The
symmetry property (\ref{symmetry1}), the relation (\ref{CtoW}) and the formula
(\ref{formulaW}) remains valid. The solution $\eta_1(\sigma)$, specified
by the boundary conditions (\ref{etadef}) with $K$ replaced
by $\frac{\pi}{2}$ is
\begin{equation}
\eta_1(\sigma)=(\omega\cos\sigma)^2F\left(1+i\frac{\omega}{2},1-i\frac{\omega}{2};
\frac{5}{2};\cos^2\sigma\right)
\end{equation}
and $\xi(\sigma)=\eta(-\sigma)$. It follows from (\ref{formulaW}), the formula (\ref{hypergeo1})
-(\ref{hypergeo3}) for hypergeometric functions that
\begin{eqnarray}
W[\eta_1,\xi_1]&=&-2\lim_{\epsilon\to
0^+}\eta_1(-\epsilon)\eta_1^\prime(-\epsilon)
=-\frac{\omega^4(\omega^2+4)}{5}\frac{\Gamma\left(\frac{5}{2}\right)
\Gamma\left(\frac{7}{2}\right)\Gamma^2\left(\frac{1}{2}\right)}
{\Gamma\left(\frac{3+i\omega}{2}\right)\Gamma\left(2+i\frac{\omega}{2}\right)
\Gamma\left(\frac{3-i\omega}{2}\right)\Gamma\left(2-i\frac{\omega}{2}\right)}\\
&=&-\frac{9\omega^3}{\omega^2+1}\sinh\pi\omega
\end{eqnarray}
Divided by $-3\omega$ we end up with (\ref{soluble12}). The nonzero
component of the spin connection corresponding to the metric
(\ref{ads2}) is $\omega_\tau^{01}=\tan\sigma$ and the counterpart of
eq(\ref{potential+}) with the upper sign reads
\begin{equation}
\frac{d^2\phi}{d\sigma^2}-\left(\omega^2+\frac{1}{1+\sin\sigma}\right)\phi=0.
\label{hateqsol}
\end{equation}
The equation (\ref{hateqsol}) can be reduced to a hypergeometric
equation and the solutions satisfying the boundary conditions
(\ref{etadef}), (\ref{cond1}) and (\ref{cond2}) (with $K$ replaced by $\frac{\pi}{2}$) read
\begin{equation}
\eta_+=2\omega^2(1+\sin\sigma)F\left(1+i\omega,1-i\omega;
\frac{5}{2};\frac{1+\sin\sigma}{2}\right)
\end{equation}
and
\begin{equation}
\xi_+=\frac{1}{\sqrt{2}}(1+\sin\sigma)\sqrt{1-\sin\sigma}
F\left(\frac{3}{2}+i\omega,\frac{3}{2}-i\omega;\frac{3}{2};\frac{1-\sin\sigma}{2}\right).
\end{equation}
Their Wronskian
\begin{eqnarray}
W[\eta_+,\xi_+]&=&\eta_+\left(\frac{\pi}{2}\right)\xi^\prime_+\left(\frac{\pi}{2}\right)
=-4\omega^2F\left(1+i\omega,1-i\omega;\frac{5}{2};1\right)\nonumber\\
&=&-4\omega^2\frac{\Gamma\left(\frac{5}{2}\right)\Gamma\left(\frac{1}{2}\right)}
{\Gamma\left(\frac{3}{2}+i\omega\right)\Gamma\left(\frac{3}{2}-i\omega\right)}
=\frac{3\omega^2}{\omega^2+\frac{1}{4}}\cosh\pi\omega.
\label{Csol}
\end{eqnarray}
The eq.(\ref{soluble3}) follows then from (\ref{Csol}) and (\ref{alternative}).



\begin{thebibliography}{0}
\bibitem{fn1}
To see the cancellation for the single line case, we write the
effective action (\ref{straight}) in the same form as
(\ref{parallel}), keeping in mind that $R=-2$ here.
\end{thebibliography}
\end{document}